\documentclass[pdflatex]{article}
\usepackage{graphicx}
\usepackage[utf8]{inputenc}
\usepackage{xcolor}
\usepackage{graphicx}
\usepackage{pifont}
\usepackage{amsmath,float,amssymb}

\graphicspath{{Figures/}}

\newcommand{\beginsupplement}{%
      \setcounter{section}{0}
        \renewcommand{\thesection}{S\arabic{section}}%
        \setcounter{table}{0}
        \renewcommand{\thetable}{S\arabic{table}}%
        \setcounter{figure}{0}
        \renewcommand{\thefigure}{S\arabic{figure}}%
        \setcounter{page}{1}
        \onecolumn
     }

\begin{document}

\title{The Physics of Sustainability:\\
Material and Power Constraints for the Long Term}

\author{Jos\'e Halloy$^{a,1}$, Petros Chatzimpiros$^{1}$, Fran\c{c}ois Graner$^{2}$, Thomas Gregor$^{3,4}$}

\maketitle

\noindent
$^{1}$ {Universit\'e Paris Cit\'e, CNRS, LIED UMR 8236, F-75006 Paris, France}\\
$^{2}$ {Universit\'e Paris Cit\'e, CNRS, MSC UMR 7057, F-75006 Paris, France}\\
$^{3}$ {Joseph Henry Laboratory of Physics  
  \& Lewis-Sigler Institute for Integrative Genomics,
  Princeton University, Princeton, NJ 08544, USA}\\
$^{4}$ {Department of Stem Cell and Developmental Biology, CNRS UMR3738 Paris Cité, 
 Institut Pasteur, 25 rue du Docteur Roux, 75015 Paris, France}\\
$^{a}$ jose.halloy@u-paris.fr

\medskip
\medskip
\medskip
\medskip
\medskip

\begin{abstract}
Much of today’s sustainability discourse emphasizes efficiency, clean technologies, and smart systems, but largely underestimates fundamental physical constraints relating to energy-matter interactions. These constraints stem from the fact that Earth is a materially closed yet energetically open system, driven by the sustained but low power-density flux of solar radiation. This Perspective reframes sustainability within these axiomatic limits, integrating relevant timescales and orders of magnitude. We argue that fossil-fueled industrial metabolism is inherently incompatible with long-term viability, while post-fossil systems are surface-, materials-, and power-intensive. Long-term sustainability must therefore be defined not only by how much energy or material is used, but also by how it is used: favoring organic, carbon-based chemistry with limited reliance on purified metals, operating at low power density, and maintaining low throughput rates. Achieving this requires radical technological shifts toward life-compatible systems and biogeochemical circular processes, and, likely as a consequence, a paradigm change toward degrowth to a steady-state. These two shifts are mutually reinforcing and together provide the necessary foundation for any viable future.
\end{abstract}

\newpage
\section{Introduction}

Human civilization has never been more powerful, nor more precarious. Over the past two centuries, industrial societies have undergone unprecedented acceleration in their ability to extract, transform, and consume. Powered by fossil fuels and enabled by mineral abundance, this expansion improved productivity, lifespan, and connectivity. But it has also destabilized the very systems that made modern human development possible.

Since the Industrial Revolution, the combustion of fossil fuels has replaced human and animal labor as the dominant source of work, triggering orders-of-magnitude increases in power use, materials transformation, and waste production~\cite{smil2013harvesting,crutzen2006anthropocene}. Global biogeochemical cycles have been altered, ecosystems degraded, and extinction rates accelerated well above background levels~\cite{steffen2015,lewis2015defining,zalasiewicz2018anthropocene}. 

The scale and speed of these transformations now raise fundamental questions about the long-term viability of industrial civilization~\cite{meadows1972limits,meadows2012limits,ripple2017world}. 
Today, natural resource depletion, climate change, and biodiversity loss continue to accelerate~\cite{Rockstrom2009,steffen2015,steffen2018trajectories,Richardson2023}---despite decades of environmental awareness and policy effort. More than 10,000 scientists have issued a global warning that our current trajectory is unsustainable~\cite{ripple2019warning}.

Many proposed solutions remain narrowly framed. Climate change, energy scarcity, and pollution abatement are often treated as separate crises, each with its own technological fix. Yet these problems are not isolated: they are deeply interconnected symptoms of the same physical overshoot~\cite{krausmann2017global,haff2014technology}. Most approaches also reduce sustainability$^\ast$ (see Box 1) to a question of quantities: tons of carbon, petajoules of energy, gigatons of material, or even economic values and innovation; while often neglecting crucial factors such as rates of use (power in W), resource concentration, and regeneration timescales~\cite{hall2009eroi,Schramski2020,odum1971environment}.

This Perspective starts from a different premise: that sustainability is primarily a physical and biological challenge. Our aim is to clarify the foundational premises that any viable strategy must respect. It requires grounding in thermodynamics~\cite{Herbert2023}, Earth system science, and biology. 

We begin in Section~\ref{sec:limits} by framing the Earth's surface as a materially closed$^\ast$ and energetically open$^\ast$ system, driven primarily by the sustained but relatively low-density flux$^\ast$ (per m$^2$) of solar radiation~\cite{Georgescu1971,schramski2015human}. 
We review key physical and biological constraints: the limits of energy throughput, material concentration, and biogeochemical cycles prior to industrial intervention. 

Section~\ref{sec:current_usage} compares these constraints with contemporary patterns of human resource use, emphasizing the dramatic increases in material throughput, material concentration, and power density that define current industry.

Section~\ref{sec:paradigm_gaps} explores why prevailing industrial paradigms fall short. It examines the false promise of ``green growth'' and then highlights how increasing efficiency and renewable substitution often fail to reduce total ecological burden. We argue that the core issue lies in systemic overshoot: too much, too fast, too concentrated.

Finally, in Section~\ref{sec:building_sustainability}, we outline a dual pathway toward sustainability. The first is a multi-factor degrowth$^\ast$ transition that significantly reduces total material and power throughput~\cite{latouche2009farewell,kallis2018degrowth}. The second involves the development of life-compatible$^\ast$ technologies: systems that operate within solar power budgets, maintain real material circularity, and are compatible with biological regeneration timescales~\cite{vincent2006biomimetics,benyus1997biomimicry}. Together, these shifts mark a qualitative and quantitative move toward ecological realism. A sustainable future is possible, but only if humans start from the right biophysical constraints.

{\it \small 
\section*{Box 1: Key concepts and definitions}
\addcontentsline{toc}{section}{Box 1: Key concepts and definitions}
~

\textbf{Ecosphere:} The thin shell at the Earth's surface where life exists, overlapping with the biosphere but emphasizing physical interfaces and fluxes.

\textbf{Closed vs open systems:} The ecosphere is materially closed (no large-scale import/export of matter) but energetically open (solar input, infrared output). 

\textbf{Stock vs flux:} On civilization timescales, a stock is a finite reserve (e.g., fossil fuels, uranium, metal ores) that can be drawn down to exhaustion, whereas a flux is a recurrent flow of energy or matter (e.g., sunlight, wind, rainfall, biomass) whose renewal is spatially and temporally bounded. Even fluxes can be depleted if exploited faster than they are renewed (over-exploitation).

\textbf{EROI vs PDROI:} Energy Return on Investment (EROI) quantifies energy gain vs energy spent. Power Density Return on Investment (PDROI) adds spatial and temporal constraints, addressing real-world deployment limits.

\textbf{Hard, soft or environmental limits:}
Hard limits on a resource arise from physics (e.g., material scarcity or thermodynamics). Soft limits, or ``reserves'', emerge from practical constraints (e.g., technical or geographical access difficulties).  
Environmental limits, or ``planet boundaries'', are damage levels beyond which the planet's habitability for humans decreases.

\textbf{Zombie vs life-compatible technologies:} Technologies based on high power and non-recyclable stocks persist and spread, driving both humanity and the ecosphere toward collapse---like zombies. Systems that use renewable energy fluxes at low power density, abundant or recyclable materials, and bio-sourced processes are life-compatible.

\textbf{Degrowth:} The intentional scaling down of material use and power within Earth's metabolism that qualitatively and quantitatively reshapes economies and inequalities.

\textbf{Sustainable}: A system is sustainable if, over long time scales (millennia), it maintains its own viability while keeping its impact on the ecosphere compatible with continued habitability. Sustainability requires both degrowth and life-compatible processes. 

}

\begin{figure}[t!]
(A) \hfill \hfill \hfill  (B) \hfill \hfill  ~\\  
    \includegraphics[height=0.45\columnwidth]{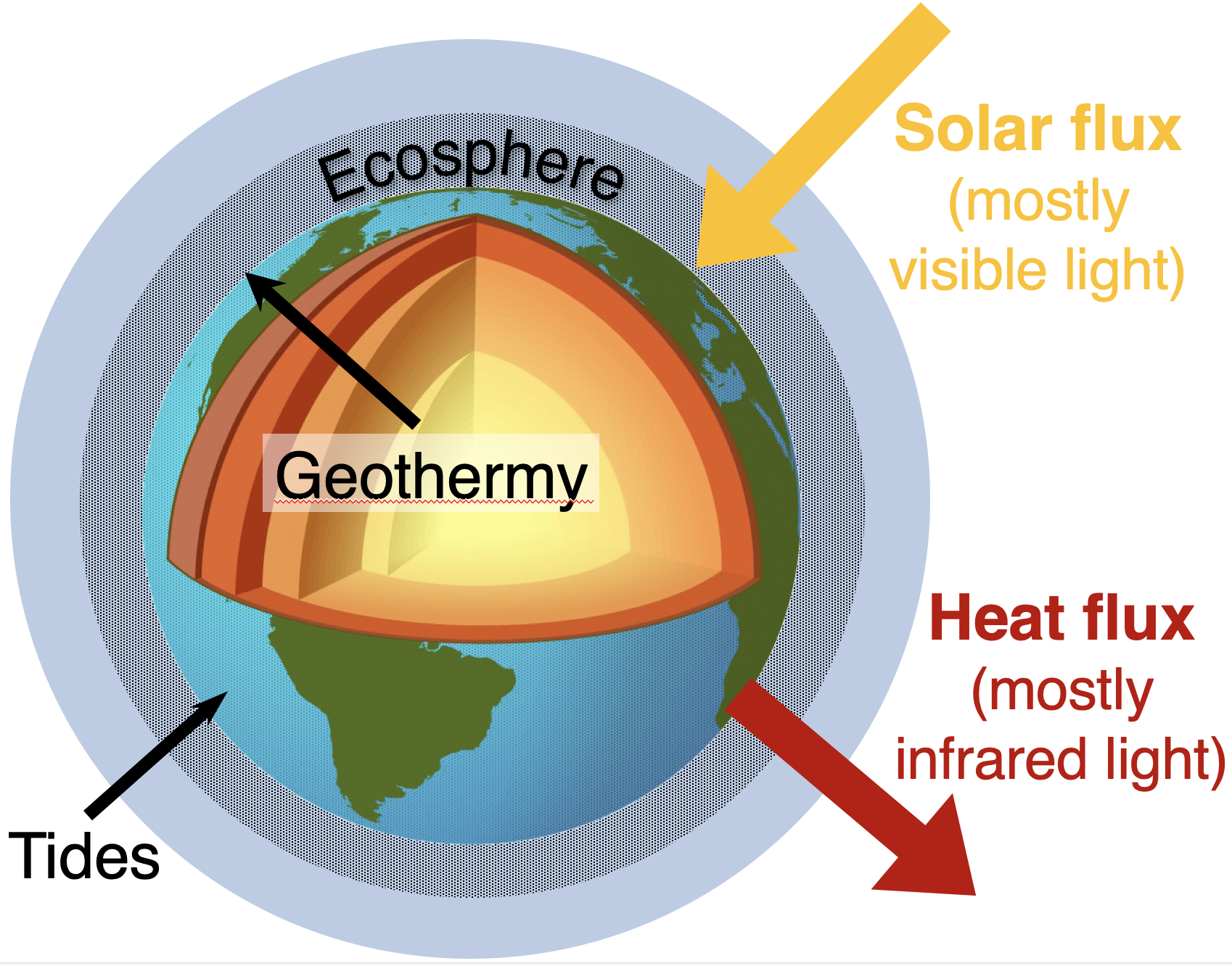}
    \includegraphics[height=0.45\columnwidth]{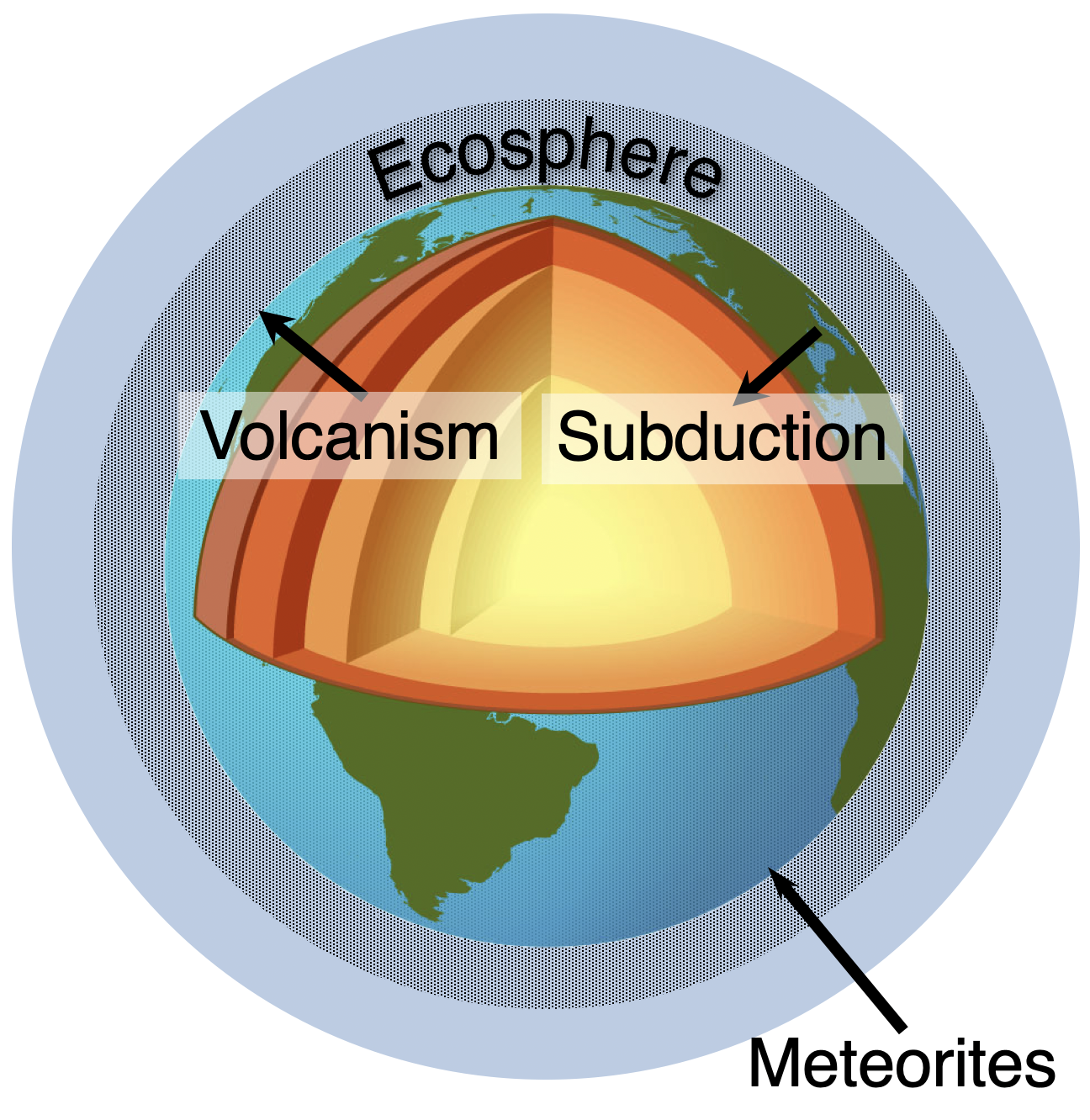}
\caption{%
\textbf{Energy and matter fluxes in the ecosphere.}
The ecosphere, the thin interface of air, water, and soil that sustains life, is shown (not to scale) at the boundary between the atmosphere and Earth’s crust. 
Only net input/output fluxes are depicted.  
\textbf{(A) Energy fluxes.} Solar radiation dominates the input ($\sim\!10^{17}$~W at the top of the atmosphere), with minor contributions from geothermal heat and tidal friction. Outgoing infrared radiation balances inputs at steady state.  
\textbf{(B) Matter fluxes.} The ecosphere’s total mass remains approximately stable. Net exchanges include meteorite infall, volcanic outgassing, and subduction; other fluxes (e.g., gas escape) are negligible at planetary scale.  
Together, these panels emphasize that Earth is materially closed but energetically open, sustained primarily by a dilute solar flux and bounded by finite matter.
}    
\label{fig:global_energy_and_matter_fluxes}
    \label{fig:Energy_fluxes_Earth_model}
    \label{fig:Matter_fluxes_Earth_model}
\end{figure}

\section{The physical constraints of the biosphere}
\label{sec:limits}

Earth harbors carbon-based life confined to a narrow zone of activity, the ecosphere$^\ast$: a thin shell only $\sim\!10^3$m thick, where air, water, and soil interact. Though this life-supporting interface accounts for less than $\sim\!10^{-3}$ of Earth's total mass or volume, it sustains all known life~\cite{schramski2015human}.

In this section, we recall key thermodynamic principles: energy and matter are conserved in quantity (they are extracted and converted, never created or destroyed) but degrade spontaneously in quality, from concentrated to diluted forms. 
We outline the biophysical constraints imposed by planetary energy and matter fluxes, and their consequences for life in the ecosphere. For clarity, each topic is treated separately, though they are deeply interconnected in practice. To emphasize core physical trends, as well as for clarity and comparability, we express quantities as orders of magnitude only, in SI units.

\subsection{Energy flows and planetary balance}
\label{subsec:energy_prehuman}

Energy quantifies the capacity to induce change. Its unit, the joule (J), allows direct comparison across processes and forms (Table~\ref{table:Babel_units}). Power is defined here in its physical sense: the rate at which energy is delivered or transformed, measured in watts (W = J~s$^{-1}$) (Fig.~\ref{fig:Powers_orders_of_magnitude_compared}, Table~\ref{table:powers}).

Fig.~\ref{fig:Energy_fluxes_Earth_model}A provides a simplified schematic of the main energy fluxes entering and exiting the ecosphere. Solar radiation, at $\sim\!10^{17}$~W incoming at the top of the atmosphere~\cite{odum1971environment}, is the dominant input. It is nearly balanced by outgoing radiation, yielding quasi-steady-state conditions.
By contrast, geothermal input (from internal planetary heat and radioactive decay) is much smaller at $\sim\!10^{13}$~W. 
Tidal contributions, driven by gravitational interactions with the Moon and Sun, are even smaller, at $\sim\!10^{12}$~W. 

The total amount of energy is conserved (first principle of thermodynamics), but its usable form degrades (second principle). Since energy is neither created nor destroyed, nearly all incoming energy is eventually dissipated as infrared radiation emitted back into space, at $\sim\!10^{17}$~W. A small imbalance ($10^{14}$-10$^{15}$~W) contributes to changes in ecosphere temperature and stored energy, particularly in biomass.

\subsection{Material scarcity and thermodynamic irreversibility}
\label{subsec:material_prehuman}

The elements that make up the ecosphere are, like energy, essentially conserved. For example, the carbon, nitrogen, phosphorus, and even gold atoms present today are the same as those formed with the Earth itself. In terms of mass, the total inventory of each natural element remains nearly constant over timescales relevant to biological and societal activity.

Fig.~\ref{fig:Matter_fluxes_Earth_model}B shows the material fluxes at the boundary of the ecosphere. 
Marginal gains occur through meteorite deposition ($\sim$~1~kg~s$^{-1}$),  cosmic rays, and solar wind (each much less than $\sim\!10^{-7}$~kg~s$^{-1}$).
Losses arise from gas escape into space and nuclear mass loss from natural radioactivity (each $\sim\!10^{-7}$~kg~s$^{-1}$) and are also marginal. 
Volcanism and subduction induce large but internally balanced flows ($\sim\!10^4$~kg~s$^{-1}$), which do not alter the net mass of the $\sim\!10^3$~m-thick ecosphere shell ($\sim\!10^{21}$~kg).
During the time since Earth formation ($\sim\!10^{17}$~s), the net change of mass of the ecosphere is negligible.

Chemical compounds tend to degrade in chemical potential (i.e., usability) through processes such as oxidation, dispersion, or dilution. Such transformations do not alter the total mass of elements. They can occur spontaneously, sometimes even releasing energy, but reversing them requires external energy input.

\begin{figure}[t!]
    \centering
  (A) \hfill  ~ \\
   \includegraphics[width=0.6\columnwidth]{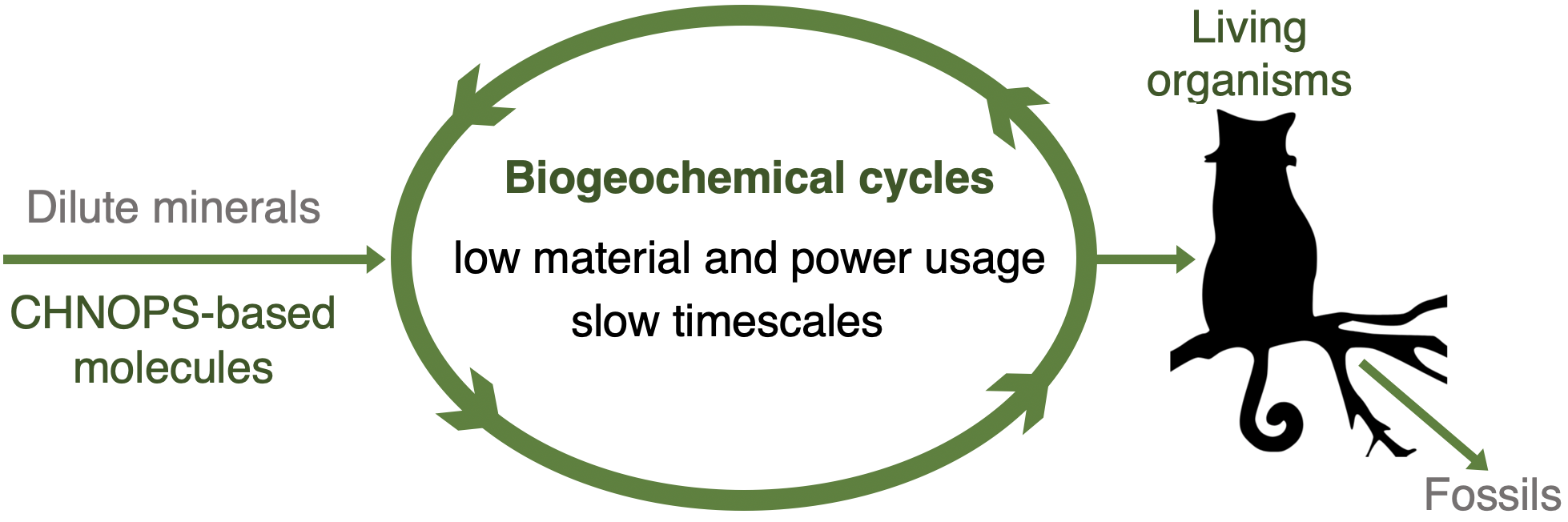}
    \\
 (B) \hfill  ~ \\ 
   \includegraphics[width=0.6\columnwidth]{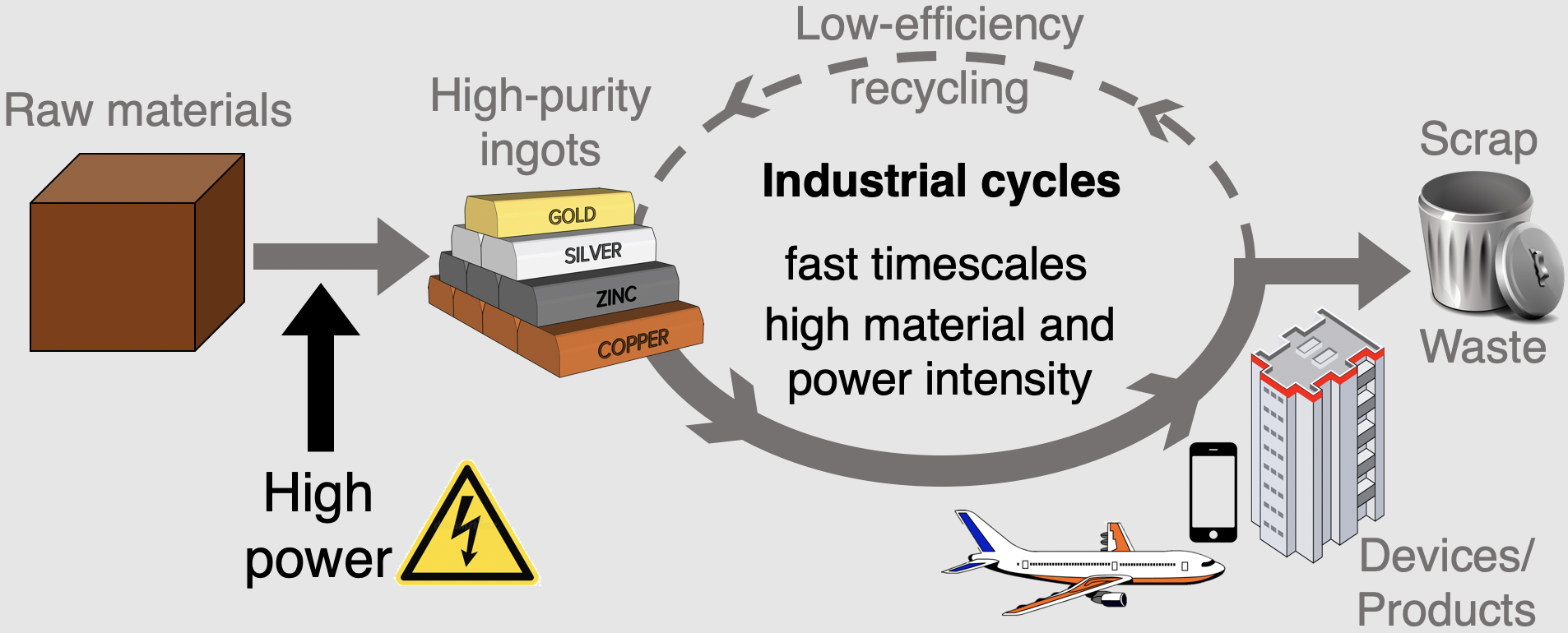}
     \\
   (C) \hfill  ~ \\
   \includegraphics[width=0.6\columnwidth]{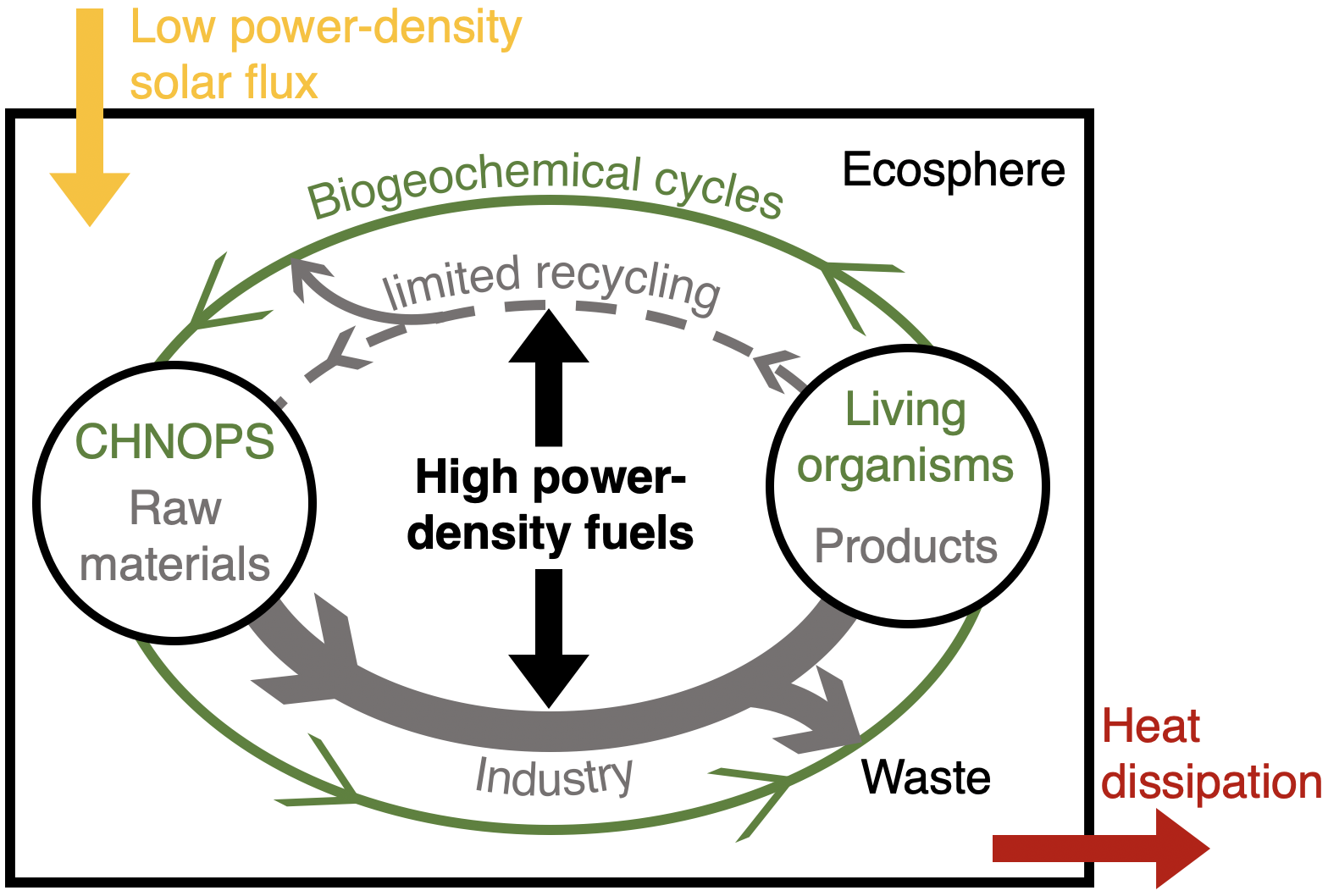}
\caption{%
{\bf Contrasting metabolisms of life and industry.}
Arrows represent material and energy fluxes.  
(A) {\it Life metabolism.} 
Biogeochemical cycles at $\sim\!10^7$~kg~s$^{-1}$ (green thick loop) recycle carbon (C), hydrogen (H), nitrogen (N), and oxygen (O), while phosphorus (P) and sulfur (S) are supplied from rocks at lower rates ($\sim\!10^3$~kg~s$^{-1}$, thin arrow)~\cite{smil2000phosphorus,holser1989sulfur}.  
Fossil accumulation (black arrow) occurred over a geological timescale ($\sim\!10^{15}$~s) via photosynthesis, leading to fossil carbon storage in the form of oil, coal, gas, and peat at rates of $\sim\!10$~kg~s$^{-1}$ and $\sim\!10^9$~W. (B) {\it Industrial metabolism.} 
Industrial systems extract concentrated raw materials at ~$\sim\!10^6$~kg~s$^{-1}$ and convert them into high-purity inputs and complex devices via high-power processes ($\sim\!10^{13}$~W, black arrow). Flows (thick grey arrows) are largely linear, with low recycling ($\sim\!10^5$~kg~s$^{-1}$, thin grey arrows), high waste, and short timescales.
(C) {\it Superimposed systems.} 
Biological and industrial metabolisms coexist within a materially closed but energetically open Earth. Life cycles remain solar-powered and slow, while industry imposes rapid, high-volume flows that disrupt natural cycles. Together, the panels illustrate the metabolic mismatch between life’s circular flows and industry’s linear, stock-based throughput.
}
    \label{fig:from_ecosphere_to_human_usages}
    \label{fig:Sketches_life_diluted_usage_and_recycling}
    \label{fig:Sketches_industry_purification_then_dilution}
    \label{fig:Recycling_good_life_poor_industry}
     \label{fig:metabolism}
\end{figure}

\subsection{Life's elemental metabolic constraints}
\label{subsec:life_usages}

Life operates within a narrow band of physical conditions: low temperatures, low pressures, low energy fluxes, and low material cycles (Fig.~\ref{fig:Sketches_life_diluted_usage_and_recycling}A). 
Solar flux drives essential planetary processes, including water evaporation ($\sim\!10^{16}$~W), atmospheric and oceanic circulation ($\sim\!10^{15}$~W), and photosynthesis ($\sim\!10^{14}$~W)~\cite{smil2017energy,odum1971environment}. 

Organic matter is primarily composed of six key elements: carbon (C), hydrogen (H), nitrogen (N), oxygen (O), phosphorus (P), and sulfur (S), collectively referred to as CHNOPS. 
Of these, C, H, O, and N account for roughly 95\% of biomass by mass and participate in active regenerating cycles across air, water, and soil. 
These cycles operate at low throughput and have remained stable over billions of years.

Carbon cycles through photosynthesis and respiration at rates near $\sim\!10^7$~kg~s$^{-1}$, with volcanism contributing an additional  $\sim\!10^4$~kg~s$^{-1}$~\cite{schramski2015human}. 
Hydrogen and oxygen are tightly coupled through water cycling and the photosynthesis-respiration loop. These cycles involve evaporation, condensation, freezing, melting, sublimation, infiltration, and both surface and subsurface flows. A unique feature of H is its partial escape from the upper atmosphere into space. 

Nitrogen is abundant in the atmosphere ($\sim\!10^{18}$~kg), but its biological fixation in natural terrestrial and aquatic ecosystems occurs at relatively low rates ($10^3$-10$^4$~kg~s$^{-1}$~\cite{Battye2017}). Nitrogen fixed by specific micro-organisms enters food webs, while a substantial fraction is dispersed through leaching and volatilization. Volcanic activity contributes an additional $\sim\!10$~kg~s$^{-1}$ of nitrogen to the lithosphere.
 
Sulfur cycles are constrained by the short atmospheric lifetime of its gaseous compounds ($\sim\!10^5$~s)~\cite{andreae1990ocean}, limiting its spatial redistribution.  

Phosphorus, unlike the other CHNOPS elements, has no significant gaseous phase. It cycles slowly ($\sim\!10^4$~kg~s$^{-1}$), with sedimentary losses ($\sim\!10^3$~kg~s$^{-1}$) balanced by tectonic uplift~\cite{smil2000phosphorus}. Its availability is tightly controlled by geological processes (e.g., rock erosion), making it a long-term limiting factor for biomass production~\cite{smil2013harvesting,schramski2015human}.

In addition to CHNOPS, life depends on trace metals such as iron, copper, cobalt, nickel, molybdenum, and zinc, typically present at low proportions (a millionth to a billionth of a percent, $10^{-11}$-10$^{-8}$). 
These biologically essential elements are acquired through membrane-bound transporters and other highly specific enzymes that minimize energy costs, enabling their efficient use at low concentrations.

\section{Human usage of energy and matter resources}
\label{sec:current_usage}

This section examines how human activity unfolds within the narrow biophysical constraints of the ecosphere. In particular, industrial processes rely on fossil fuels, which provide high power density, to extract and transform raw materials into high-purity inputs for technological systems. This industrial metabolism forms a distinct “technosphere”~\cite{haff2014technology}: the portion of the Earth system transformed by human processes, with its own energetic and material flows. We then consider the implications for resource depletion.
  
\subsection{High power density in the fossil fuel era}
\label{sec:current_energy}

Human usage of energy, commonly described as ``production'' or ``consumption'',  involves harvesting a pre-existing energy form and converting it into another that better suits specific needs.
Renewable sources, such as geothermal energy, tides, and sunlight-derived forms (e.g., river flow, solar heat, wind, ocean thermal gradients, biomass, food, photovoltaic electricity) all deliver low power density expressed in W~m$^{-2}$.

By contrast, chemical energy stored in hydrocarbons (natural gas, coal, oil, peat) or nuclear energy stored in uranium provides high power densities (in J~kg$^{-1}$), enabling concentrated, on-demand energy release.
Combustion engines operate at approximately $\sim\!10^3$~W~kg$^{-1}$, several orders of magnitude above the metabolic power density of biological organisms ($\sim\!1$~W~kg$^{-1}$)~\cite{smil2017energy}. 
This decoupling of power from biophysical limits enabled the Industrial Revolution: it allowed for fast, high-throughput material and energy flows without relying on circularity.
A single fossil-fuel power station can deliver $\sim\!10^9$~W, the same output as $10^7$-10$^8$~m$^2$ of solar panels.  

Power (in W), rather than energy (in J), is what humans primarily perceive and interact with in everyday actions (Fig.~\ref{fig:Powers_orders_of_magnitude_compared}). Schematically, power determines what is possible and drives differences in the organization of society. For instance, moving a heavy object horizontally, as in the construction of megalithic dolmens, the Great Wall of China, or the Egyptian pyramids, requires little power but extended time. In contrast, vertical movement (e.g., by helicopter or rocket) demands high power over short durations. In general, the achievable power depends on the conversion time of a given energy quantity. During the Industrial Revolution, humans progressively shortened conversion times. A quantity of $10^7$~J (equivalent to 250~g of gasoline) yields an average power of 100~W (the metabolic human rate) if converted in $10^5$~s (24h), 1~kW (a kitchen appliance) in $10^4$~s (2.5 hours), 100~kW (a car) in $10^2$~s and 1~GW (a rocket engine) in $10^{-2}$~s (Fig.~\ref{fig:Powers_orders_of_magnitude_compared}). 

Today, the average global power use of machines is on the order of $\sim\!10^{13}$~W, with approximately 80\% supplied by fossil fuels. For comparison, the biosphere absorbs $\sim\!10^{16}$~W of solar radiation (60\% on land, 40\% in oceans), which drives photosynthesis and generates $\sim\!10^{14}$~W in plant biomass~\cite{IPCC2013}, of which agriculture and forestry appropriate $\sim\!10^{13}$~W~\cite{Sassoon2009} (see Section~\ref{sec:sassoon_figures}). The metabolic power rate of the global population ($\sim\!10^{10}$ people), at $\sim\!10^2$~W per human, is $\sim\!10^{12}$~W. By comparison, individual machines operate at power levels several tens of times greater than the human body.

\subsection{Industrial material use: from ore to complex alloys}
\label{sec:current_material}

Compared to pre-industrial societies, modern human activity has undergone a profound quantitative transformation. Today, up to 80 of the 92 naturally occurring chemical elements are used in industrial applications, with many increasing exponentially in extraction rate~\cite{smil2016making} (Fig.~\ref{fig: Material_extraction_vs_year}). Total anthropogenic material flows now exceed $\sim\!10^6$~kg~s$^{-1}$~\cite{Krausmann2018,Schandl2024}, dwarfing natural inward and outward fluxes.

Between 1900 and 2015, humans released approximately $10^{15}$~kg of solid, liquid, and gaseous waste into the environment, with one quarter of that mass deposited in just the last decade~\cite{Krausmann2018}. In practice, material recycling rates in the European Union remain low: about 3\% for fossil fuel–derived materials such as plastics, 10\% for biomass-derived materials like textiles and construction wood, 24\% for metals, and 14\% for non-metallic minerals, with an overall average of 12\%~\cite{EEA2021,Eurostat2025}. On average, each human is now associated with the weekly production of an anthropogenic mass exceeding their own body weight (Fig.~\ref{fig:Recycling_good_life_poor_industry}B). As of 2020, the total anthropogenic mass has surpassed the Earth’s total living biomass ($\sim\!10^{15}$~kg)~\cite{elhacham2020global}. Human activity also amplifies the carbon cycle, contributing to surface warming far more than direct heat release from energy transformation.

The transformation has also been qualitative. Extracted elements must often be purified to extremely high levels—for instance, electronic-grade silicon requires impurities below 1 part per billion, before being combined into complex mixtures used in batteries, electronics, infrastructure, and machinery. 
 
Material extraction and refinement are increasingly energy-intensive. To obtain just 1~kg of copper from ore containing 1\% Cu,  10$^2$~kg of rock must be mined, crushed, and processed. Rare earths and platinum-group metals are even more demanding, both energetically and logistically. Once extracted, materials are transformed into alloys, composites, and devices, largely through unidirectional processes. They move from geological dispersion to engineered complexity and eventually to waste (Fig.~\ref{fig:Sketches_industry_purification_then_dilution}B). Although some high-volume materials like aluminum are recycled, recovery of most specialized elements is limited by contamination, thermodynamic barriers, or technological complexity (e.g., up to 20 elements in a single steel alloy, many in trace quantities)~\cite{graedel2011recycling}, and recycling remains energy-intensive~\cite{zepf2014materials}. In general, the more diverse and composite the parts of a device, the more difficult it is to recycle.

\subsection{Stocks vs fluxes: physical limits and practical constraints}

Resources can be classified into two main categories. Stocks$^\ast$ are finite accumulations, such as fossil fuels and mineral ores, that allow concentrated and on-demand extraction but are ultimately exhaustible. Fluxes, such as solar radiation, wind, or rainfall, are recurrent flows. While often renewable, they are typically diffuse, variable, and distributed in space and time.

Some resources combine both stock and flux properties. In these cases, whether a resource is exploited as a stock or a flux depends on the ratio between its renewal rate ($R_r$) and human consumption rate ($R_c$). For example, humans consume freshwater at a rate of $\sim\!10^5$~kg~s$^{-1}$, while global rainfall renews it at $\sim\!10^{10}$~kg~s$^{-1}$, about one-third of which falls on land. Since $R_r / R_c \sim\!10^5$, water is globally exploited as a flux. However, in specific regions where deep aquifers are depleted faster than they recharge ($R_r \ll R_c$), it is effectively used as a stock, leading to overdrafting.

``Hard$^\ast$ limits'' on resource use are imposed by the laws of physics. For a stock, it is the total amount of the resource in the ecosphere. For a flux, it is the maximum rate at which it is available, expressed in kg~s$^{-1}$ for matter, or W for energy. ``Soft$^\ast$ limits'', more relevant to daily life, arise when harvesting becomes inefficient or damaging relative to benefit, due to technological, ecological, or economic constraints.

For energy stocks, the soft limit is often expressed through the ``energy return on investment'' (EROI$^\ast$), the ratio of energy harvested to invested energy. It is dimensionless, and if it falls below 1, the return becomes unfavorable. A complementary indicator is the ``power density return on investment'' (PDROI$^\ast$), defined as the ratio of peak power harvested per unit area to the power per unit area required for manufacturing and distribution. For example, fossil coal and charcoal have similar energy content per unit mass ($\sim\!10^7$~J~kg$^{-1}$), yet their production power densities differ by three orders of magnitude: per unit surface, even a low-grade lignite mine produces on the order of $10^{-5}$~kg~s$^{-1}$~m$^{-2}$, whereas a forest yields only about $10^{-8}$~kg~s$^{-1}$~m$^{-2}$ as charcoal.

For matter, no equivalent dimensionless indicator exists. Ratios such as the energy return on invested matter (J~kg$^{-1}$) or the matter return on invested energy (kg~J$^{-1}$) conflate two non-exchangeable entities—energy and materials—and provide no absolute threshold for defining a soft limit. Even the ratio of matter recovered (kg) to matter invested (kg) is ill-defined: one kilogram of lithium is not equivalent to one kilogram of iron, and the 92 naturally occurring elements cannot be meaningfully compared on a mass basis alone. Thus, matter lacks a universal return-on-investment metric, despite its centrality to sustainability. Proposals such as deep-sea mining, polar colonization, or asteroid extraction are often suggested as ways to extend such limits, but their feasibility remains speculative and unproven in practice. This absence reinforces the need to frame resources according to their PDROI and to their basic distinction as stocks or fluxes.

\begin{figure}[t!]
\begin{center}
   (A) \hfill ~\\ 
   \includegraphics[width=0.55\columnwidth]{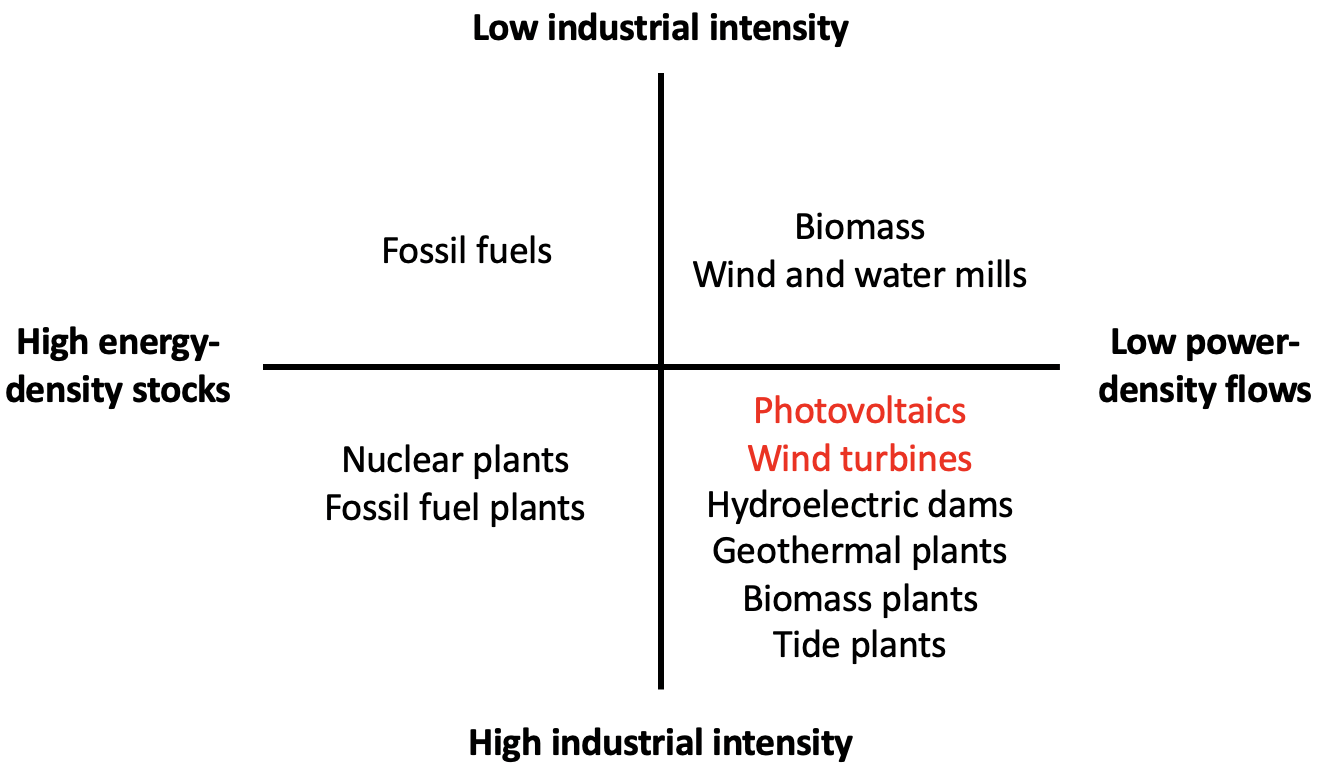}    
   \\
   \vspace{1cm}
   (B) \hfill ~\\ 
   \includegraphics[width=0.55\columnwidth]{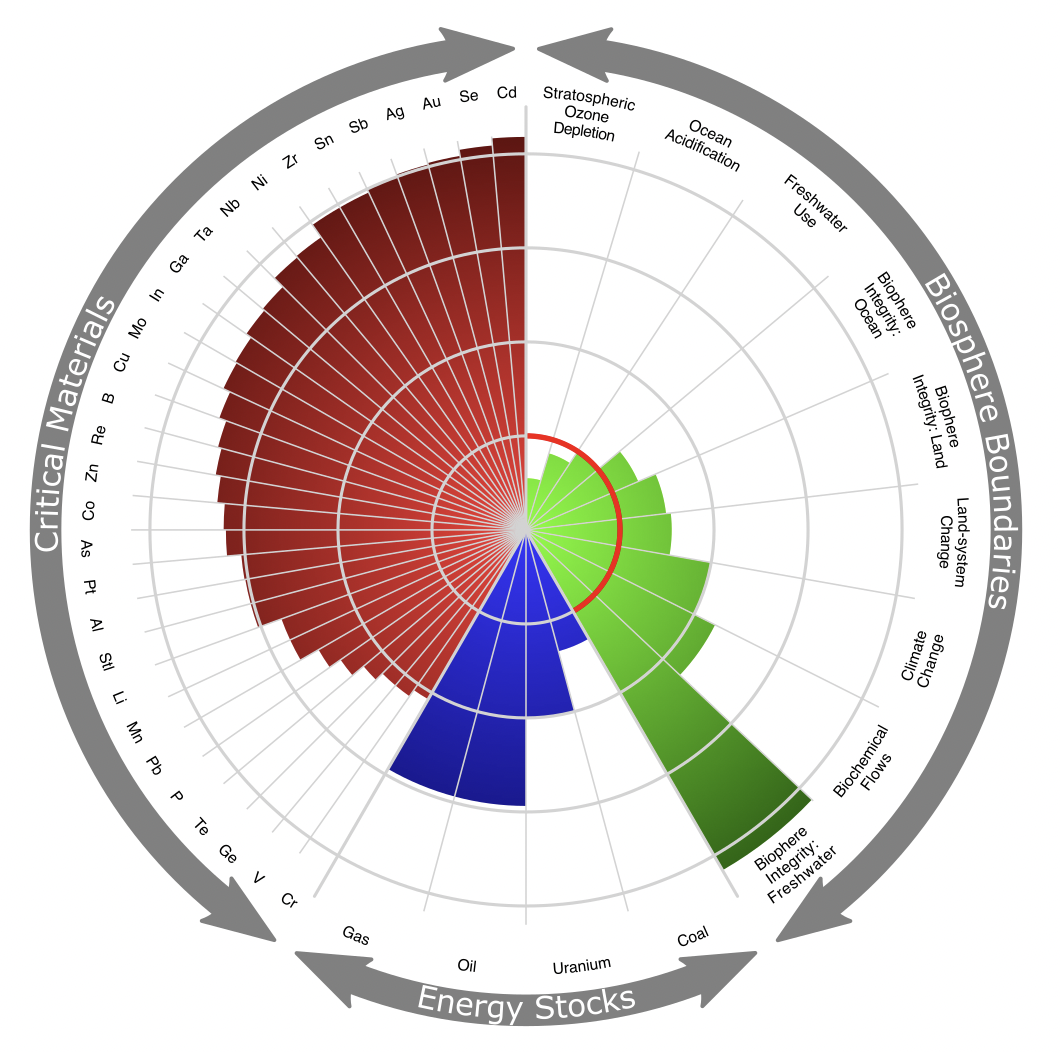} 
\end{center}
\caption{%
\textbf{Current human activity is not sustainable.}
\textbf{(A)} Common energy sources and their corresponding converters, plotted against two axes: the vertical axis shows peak power required for manufacturing and maintenance; the horizontal axis indicates the type of energy being converted. Technologies that rely on scarce materials are highlighted in red.  
\textbf{(B)} Global picture of criticality: depletion of material stocks, exhaustion of energy reserves, and violation of biosphere boundaries. Feedbacks are shown as grey arrows. Each radial sector maps a resource or boundary; a larger radius indicates a more critical condition. A timescale of 50 years enables comparison of stocks and flows at the scale of two human generations. Adapted 
from~\cite{Rockstrom2009,steffen2015,steffen2018trajectories,Richardson2023}.
\textit{Materials (left, red):} Years projected until peak extraction.  
\textit{Energies (bottom, blue):} Fossil fuel reserves plotted on the same timescale.  
\textit{Biosphere (right, green):} Boundaries updated from~\cite{steffen2015}; red circular arc marks the planetary threshold.
Together, these panels highlight the unsustainable trajectory of current human activity, marked by resource depletion and transgression of planetary boundaries.
}
\label{fig:sustainability_gaps}
\label{fig:Energy_and_embodied_power}
\label{fig:Rosetta_of_three_planetary_boundary_types}
    \label{fig:paradigm}
   \end{figure}

\section{The metabolic mismatch: Why prevailing industrial paradigms fail}
\label{sec:paradigm_gaps}

This section explores the fundamental incompatibility between industrial systems and the regenerative limits of the biosphere. Building on the couplings between power, material flow, and sustainability thresholds, we contrast the logics of biological and industrial metabolisms, assess why current activity exceeds planetary boundaries, and explain why techno-centric fixes such as ``green growth'' fail to achieve sustainability.

\subsection{Metabolic conflict: Biosphere versus industry}
\label{sec:two_metabolisms}

Fig.~\ref{fig:metabolism}C illustrates the metabolic mismatch between Earth’s biosphere and industrial civilization~\cite{crutzen2006anthropocene,steffen2015}. The biosphere operates as a closed-loop, low-throughput system powered by diffuse but steady solar input ($\sim\!10^{17}$~W), of which only $\sim\!10^{15}$~W is absorbed by the ecosphere. Material cycles are maintained through slow, multiscale biogeochemical recycling of CHNOPS elements, sustaining life over billions of years~\cite{odum1971environment}.

In contrast, industrial metabolism is linear, extractive, and powered by concentrated energy stocks. It mines high-purity mineral resources, resulting in global anthropogenic material flows of $\sim\!10^6$~kg~s$^{-1}$~\cite{krausmann2017global}, processed using $\sim\!10^{13}$~W, still about $\sim\!80\%$ fossil-derived~\cite{IEA2021}. These flows generate large non-recyclable waste~\cite{Sassoon2009} and significantly disrupt biogeochemical cycles, particularly through the accumulation of carbon and nitrogen in the biosphere~\cite{Rockstrom2009,Galloway2021}.

Technologies based on high power and non-recyclable materials bring humanity and the ecosphere to a dead end, even before accounting for their ecological impacts. Yet they persist and proliferate as ``zombie$^\ast$ technologies'': systems that continue operating despite being incompatible with long-term viability. They are characterized by five features: (i) dependence on energy stocks, (ii) dependence on material stocks, (iii) power-density mismatch, (iv) overshoot of biogeochemical cycles and biological constraints, (v) rapid senescence with poor circularity~\cite{halloychp11}. 

Fig.~\ref{fig:Energy_and_embodied_power}A schematizes this mismatch: low-tech systems (top) operate on renewable flows with modest infrastructure, while high-tech systems (bottom) rely on concentrated stocks and energy-intensive processes. 

A key example of such zombification is industrial agriculture. Although agriculture is fundamentally a CHNOPS ``technology'', industrialization has created unprecedented dependence on nonrenewable resources. This dependence has enabled a tenfold increase in population size since 1800, and has also driven a dramatic rise in animal protein consumption. Under current land use and diets, the feeding capacity of global agriculture is estimated at roughly 13 billion people with industrial fertilization, and between 4 and 12 billion with organic fertilization depending on fixation rates and nitrogen-use efficiency~\cite{Chatzimpiros2023}. The feeding capacity without fossil fuels remains an open research question and is likely much lower. 

Industrial nitrogen fixation through the Haber-Bosch process has been pivotal in raising productivity~\cite{Smil2001,Galloway2021}. Human-applied nitrogen fluxes ($\sim\!10^3$–$10^4$~kg~s$^{-1}$) across the $\sim\!10^{13}$~m$^2$ of global cropland (grasslands receive little fertilizer) now match the total reactive nitrogen fixation in natural ecosystems~\cite{Battye2017,Sutton2013}. Global nitrogen-use efficiency is currently about 20\%, meaning that for every unit of nitrogen retained in food, roughly four units are lost to the environment~\cite{Chatzimpiros2023}. The total fossil power used to produce and deliver food (including agrochemicals, mechanization, shipping, processing, refrigeration) exceeds the edible power delivered as food ($\sim\!10^{12}$~W) by a factor of three to six~\cite{Schramski2020}.

Phosphorus presents a particular constraint: harvesting depletes soils, and because it has no gaseous phase, there is no natural return path. As a result, continued reliance on phosphate fertilizers draws down finite rock reserves. This dependency highlights a broader vulnerability of industrial agriculture, which can expand production in the short term while eroding the very cycles needed for long-term sustainability. 

\subsection{Systemic overshoot and planetary boundaries}
\label{sec:overshoot_limits}

Industrial metabolism not only conflicts with biospheric cycles but also drives systemic overshoot of planetary boundaries. Every stage of this metabolism, from extraction to production, operation, and disposal, requires both energy and matter while generating waste (Fig.~\ref{fig:Recycling_good_life_poor_industry}C). Due to conservation laws and thermodynamic degradation, nearly all material and energy flows eventually dissipate into air, water, or soil. Today, the mass of anthropogenic material exceeds that of all living biomass on Earth~\cite{elhacham2020global}.

Environmental$^\ast$ limits, such as climate thresholds or biodiversity collapse, represent damage levels beyond which the planet becomes less habitable for humans. These ``planetary boundaries'' are difficult to quantify precisely. Biodiversity, in particular, is complex, dynamic, and emergent~\cite{Thiebault2024}, and cannot be adequately captured by species counts alone~\cite{Landenmark2015,baron2018biomass,Seibold2019,Regnier2015}.

Nonetheless, semi-quantitative frameworks~\cite{Rockstrom2009,steffen2015,steffen2018trajectories,Richardson2023} help illustrate the criticality of our current state. Fig.~\ref{fig:Rosetta_of_three_planetary_boundary_types}B summarizes present ecological degradation (outputs) and near-term resource constraints (inputs). Human activity is not only unsustainable, it is still accelerating (Fig.~\ref{fig:increase_of_human_usages}).

Models of resource extraction over time~\cite{hubbert1956nuclear} predict peak availability for most exploited resources. Among the 92 naturally occurring elements, ~56 are actively extracted, and most will reach peak production before 2100~\cite{heinberg2010peak,bardi2014extracted,riondet2023applicability}. The rest are expected to peak by the early 22$^\text{nd}$ century (Fig.~\ref{fig:Critical_materials_and_energy_vs_year}).

Importantly, output-side environmental constraints (climate, waste, biodiversity loss) are often more immediate and severe than input-side depletion. For instance, limiting global warming to 2$^\circ$C would require leaving 80--90\% of remaining fossil fuel reserves unburned~\cite{IPCC2024}.

\subsection{The double fallacy of ``green growth''}
\label{sec:green_growth_rebuttal}

A popular response to environmental degradation is the idea of ``green growth''~\cite{georgeson2017global}, whose realism has been widely criticized~\cite{hickelisgreengrowthpossible,Parrique2019}. From a physical standpoint, green growth is flawed and confusedly defined for two main reasons. 

At one extreme, it assumes —implicitly or explicitly— that human activity can continue to grow while all impacts on matter, energy, and the environment are simultaneously reduced. This violates the laws of thermodynamics. The quest for a perpetual motion machine was formally abandoned by the French Academy of Sciences in 1775~\cite{AcadSci1775}, and later formalized by the second law of thermodynamics, which rules out not only perpetual motion in energy but also perfect industrial recycling or ``circularity'' in materials~\cite{Giampietro2019}. 

At the other end of the discourse spectrum, green growth is reduced to a focus on greenhouse gas emissions, particularly the current global CO$_2$ flux of $\sim\!10^6$~kg~s$^{-1}$~\cite{IPCC2013}, in the fight against climate change. This so-called ``energy transition'', centered on electrification and fossil-fuel substitution, entails growth in both energy and material demand.  The infrastructures needed for this transition exemplify zombie technologies, meeting all five of the criteria defined above. They intensify demand for critical elements such as copper ($\sim\!10^3$~kg~s$^{-1}$), nickel ($\sim\!10^2$~kg~s$^{-1}$), and lithium ($\sim\!10$~kg~s$^{-1}$)~\cite{USGS2025}, with substitutes often unavailable or insufficient~\cite{graedel2015basis}. Life-cycle assessments further show that many ``green'' technologies merely displace environmental impacts across space and time~\cite{Desing_circular_economy} (see Section~\ref{sec:hurdles}). 

For instance, electric vehicles reduce tailpipe emissions but increase demand for metals, battery production, and electricity—most of which is still generated from fossil fuels~\cite{MacKay2008,goe2014identifying}. Similarly, digitization is often misleadingly framed as de-materialization. In reality, it entails substantial material and energetic costs, particularly for electronics and their infrastructures. Producing electronic-grade silicon at $\sim\!10^{-9}$ impurity levels requires $\sim\!10^{10}$~J~kg$^{-1}$~\cite{Williams2002}, and its recycling is not only thermodynamically costly but also often technologically infeasible, since these systems are not designed for recyclability~\cite{reck2012challenges}.

Taken together, these examples demonstrate the double fallacy of green growth: assuming that efficiency can outpace thermodynamic limits and that technological substitution can bypass material constraints.

\begin{figure}[t!]
    \centering
   (A) \hfill   ~\\ 
   \includegraphics[width=0.7\columnwidth]{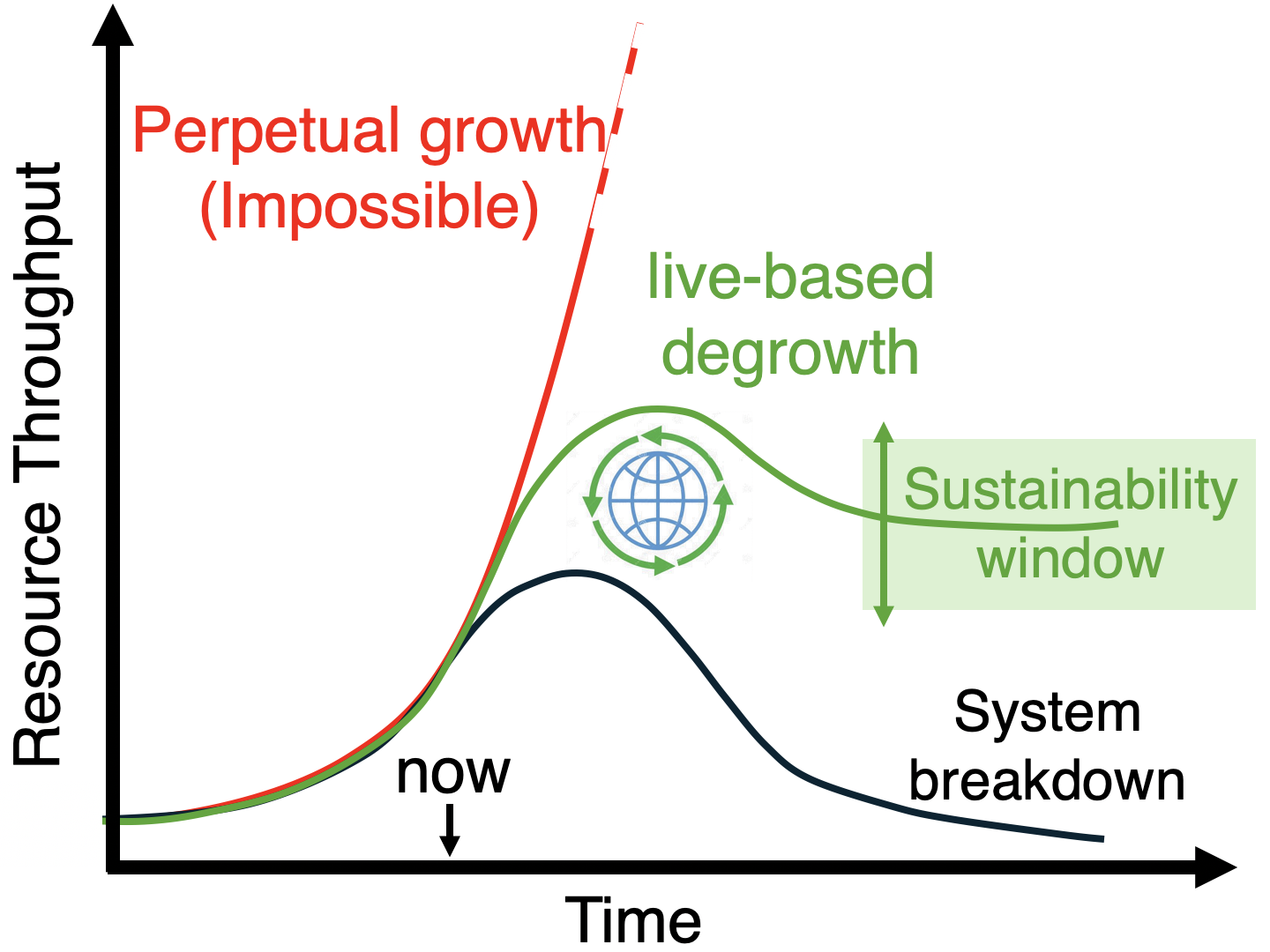}
   \\
   (B) \hfill   ~\\ 
   \includegraphics[width=0.7\columnwidth]{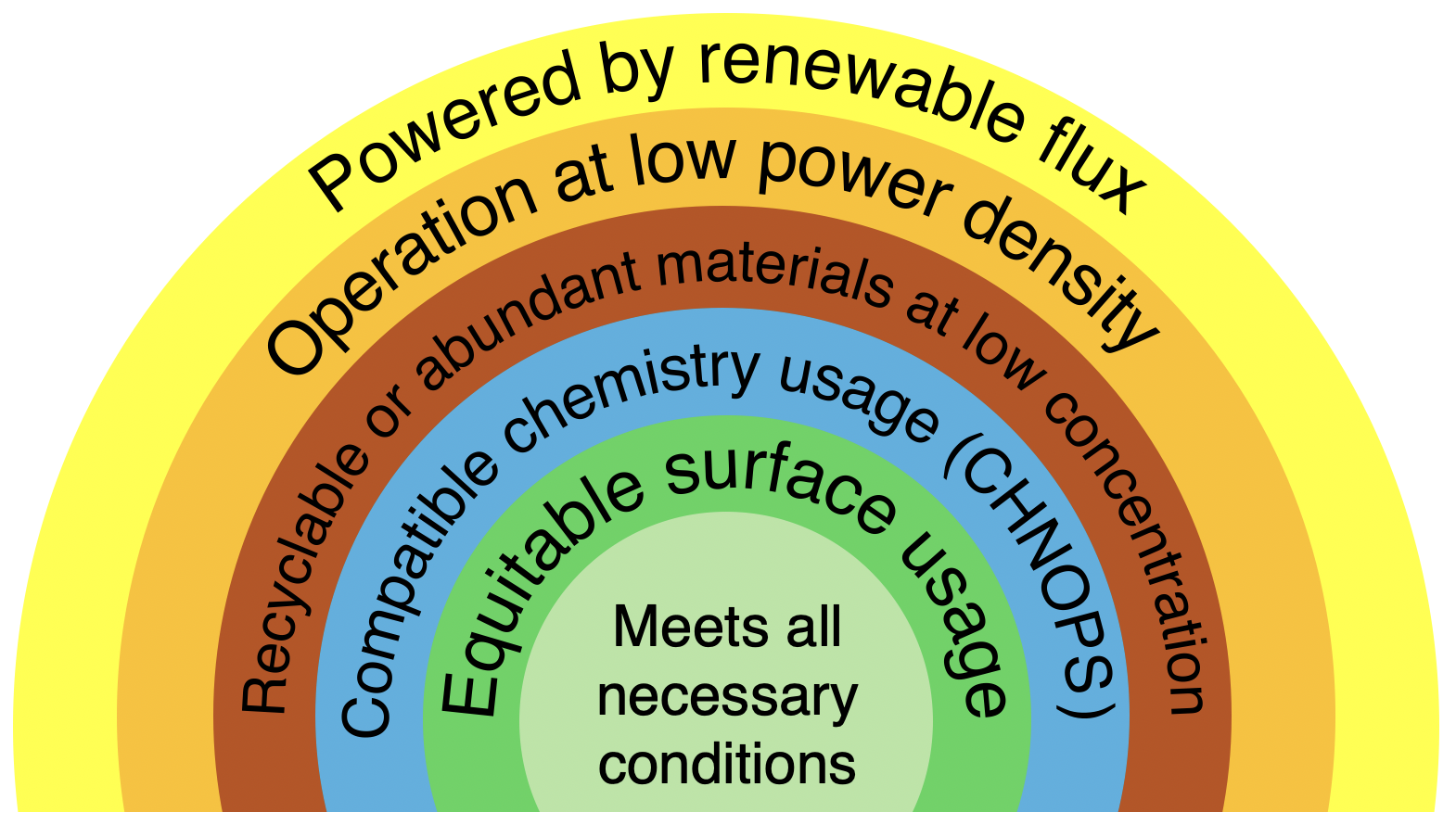}
\caption{%
\textbf{Degrowth and life-compatible technologies as dual requirements for sustainability.}  
\textbf{(A)} Schematic trajectories for industrial society. Continued throughput growth (red) is physically unsustainable; unchecked decline leads to collapse (black). A deliberate recalibration of energy and material flows, i.e., ``degrowth'' (green), offers long-term compatibility with biophysical constraints.   Adapted from~\cite{meadows1972limits}.
\textbf{(B)} Conditions for life-compatible technologies, summarized as a nested checklist (outer to inner layers): powered by renewable energy fluxes; operating at low power density; relying on recyclable or abundant metals at low concentrations; using CHNOPS-compatible chemistry; and respecting equitable land surface use. Meeting all criteria is necessary, though not always sufficient, for long-term sustainability.
This figure synthesizes the dual pathway (degrowth and life-compatible technologies) required to align human activity with planetary constraints.
}
	\label{fig:proposition}
	\label{fig:Toward_degrowth}
	\label{fig:Four_circles_energy_requirements}
    \label{fig:degrowth_figure}
     \label{fig:dual_pathways}
\end{figure}

\section{A dual path towards sustainability}
\label{sec:building_sustainability}

We outline a twofold strategy for genuine sustainability (Fig.~\ref{fig:dual_pathways}): (i) a global framework for societal transformation that deliberately reduces overall material and energy throughput while reassessing the nature of the materials used (Section~\ref{subsec:degrowth}), and (ii) an implementation based on re-engineering technological approaches to ensure compatibility with biospheric constraints (Section~\ref{subsec:life_like_technologies}). These two paths—conceptual and practical—are complementary~\cite{Kerschner2018} and deeply interdependent.

\subsection{Towards degrowth}
\label{subsec:degrowth}

As mentioned in Section~\ref{sec:limits}, the physical constraints that govern power, matter, and the environment are deeply interwoven. Addressing them requires a systems-level perspective grounded in feedback, self-organization, and emergent dynamics. The central challenge is not merely to avoid exhaustion or increase efficiency, but to resolve the structural mismatch between industrial dynamics and biospheric functioning, encompassing underlying chemistry, power, and biogeochemical cycling. 

Fig.~\ref{fig:degrowth_figure}A schematizes three scenarios. Perpetual exponential growth in energy and material throughput (red), powered by high-density fossil stocks and dependent on resource concentrations and transformation rates orders of magnitude above biospheric norms, is physically unsustainable in the long term, as discussed in Sections~\ref{sec:two_metabolisms},~\ref{sec:overshoot_limits}. At the other extreme, nothing in physics prevents collapse into zero activity and human extinction (black). Between these extremes, the only physically viable window is a degrowth trajectory (green), a transient process of intentionally scaling down material use and power within Earth's metabolism, defined in physical quantities (kg, m, s, J, W) rather than monetary terms. This path is slowed, localized, and regulated by feedback loops. It is re-coupled to biospheric regeneration and respectful of environmental boundaries. Degrowth should be understood not as an endpoint but as a process aimed at enabling a range of stable, sustainable steady states. The precise range of such sustainable steady-states is still unknown and remains an important research question.

While we maintain a strict biophysical standpoint here, we acknowledge that implementing degrowth has deep qualitative and quantitative social implications. Any systemic response to physical limits must extend to political, economic, and behavioral transformations. It demands institutional, analytical, and epistemic shifts across disciplines and sectors. Decision-making can be informed by models that integrate economics, mathematics, sociology, and information theory~\cite{briens_phd_2016}. Foundational sciences such as astronomy, geology, climatology, and evolutionary biology offer crucial perspectives for long-term thinking. Degrowth spans numerous academic domains: from physics and environmental science to philosophy, anthropology, demography, and political theory~\cite{Martin2016}.

In fact, scholarly interest in degrowth has grown markedly in recent years across disciplinary boundaries~\cite{Lehmann2022,Lewis2023,Blanchard2024}. Social scientists, too, recognize that energy and matter are conserved in quantity but degrade in quality~\cite{Georgescu1971}. Grubler~\cite{Grubler2018} and Graedel et al.~\cite{graedel2015basis} emphasize that long-term human viability depends more on absolute reductions in resource use than on efficiency or substitution. Hickel~\cite{Hickel2020} and Kallis~\cite{kallis2018degrowth} frame degrowth as a response to the physical impossibility of infinite throughput on a finite, materially closed planet. Living better for all requires deep reductions in energy and material consumption; in waste and environmental impact; in population size; in social and geographic inequality; and in the power, speed, and scale of human activity.

From a biophysical standpoint, degrowth is best understood as a thermodynamic constraint: material fluxes must return to levels that can be indefinitely sustained by solar-driven cycles, with low entropy production, long lifetimes, and minimal irreversible transformations. This implies drastically reducing high-power industrial activities, including extraction, transport, and manufacturing processes that exceed the biosphere’s renewal rates. What is required are large, absolute reductions with genuine ecological effects: beyond marginal improvements, relative metrics, or directional shifts. Degrowth must therefore be understood as a reduction in power throughput, a move away from technologies dependent on extreme material purity, and a return to chemistry aligned with life’s regenerative cycles.
 
Exponential growth has driven orders-of-magnitude increases in rates of resource consumption (Fig.~\ref{fig:increase_of_human_usages}), along with rising waste and environmental degradation. Restoring sustainability now requires reductions of comparable scale. Instead of prioritizing expansion, throughput, and obsolescence, this approach emphasizes preservation, redistribution, anticipation, maintenance, and especially reuse, which often outperforms recycling. Processes must slow to timescales that allow biological and geochemical feedback to function.

Living systems have long thrived in far-from-equilibrium dynamics, so long as they remained compatible with biophysical limits. Whether framed as ``degrowth'' or simply as physical realism, the conclusion is the same: enduring human systems must operate within the constraints life has obeyed over billions of years, and the path toward this goal must involve a deliberate descent in throughput; together with the design and adoption of technologies capable of operating indefinitely within these same biophysical limits.

\subsection{Towards life-compatible technologies}
\label{subsec:life_like_technologies}

Building on the degrowth and steady-state range framework outlined above, we turn to the complementary requirement: the development and deployment of life-compatible technologies. These must satisfy a suite of physical and ecological constraints, introducing new scientific and engineering challenges that span both natural and social domains. Viable responses, here termed ``life-compatible technologies'', fall into three broad categories~\cite{Manzini2015,Vergragt2011}:

\textit{First}, life itself demonstrates that sustainable solutions to complex constraints are not only possible but have persisted for nearly 4 billion years. The biosphere has operated through a consistent physico-chemical metabolism embedded within planetary feedback loops (Fig.~\ref{fig:Sketches_life_diluted_usage_and_recycling}A), maintaining dynamic balance through continual planetary change and biological evolution. This longevity does not stem from any intrinsic virtue of biological systems; indeed, many organisms disrupt their environments~\cite{haff2014technology}. Before the Great Oxidation Event, for example, oxygenic photosynthesis by early cyanobacteria released what was then a toxic byproduct~\cite{Lenton2011}, triggering mass extinctions and forcing much of the biosphere into anaerobic habitats~\cite{Sessions2009}.

Life also generates waste: calcium shells, hydrocarbon burial (before lignin-degrading fungi evolved), and other detritus accumulate without full recycling. Interactions between organisms, cooperative or competitive, produce fluctuations and systemic shocks, including mass extinctions. Yet across geological timescales, life as a whole exhibits remarkable resilience and compatibility with biospheric boundaries. The emergence of oxygenic photosynthesis, for example, irreversibly altered atmospheric composition and enabled the rise of aerobic metabolism~\cite{schramski2015human}.

This enduring sustainability arises from key features: closed-loop biogeochemical cycling; use of abundant elements (CHNOPS); catalytic chemistry at ambient temperature and pressure; and reliance on dilute, steady energy fluxes from sunlight and geothermal gradients. Life is slow, modular, and energetically frugal. Though individual lineages perish, those that persist are filtered by deep evolutionary compatibility with long-term environmental limits.

\textit{Second}, some existing and traditional technologies already meet multiple criteria for long-term sustainability. Examples include passive solar heating, solar thermal collectors, hydroelectric microgrids, and wind turbines; systems that operate at relatively low power densities with moderate material demands~\cite{Schumacher1973}. Likewise, long-standing techniques such as earthen architecture, root cellars for refrigeration, and composting toilets align with biospheric constraints, requiring neither rare materials nor complex infrastructure~\cite{Vergragt2011}. 

These approaches share common traits: locality, slowness, decentralization, and compatibility with rhythms of repair and reuse. Their viability is not purely technical but also cultural, sustained by social norms that value maintenance over obsolescence. However, even these ``low-tech'' solutions must be deployed with care, ensuring that material sourcing and ecological impacts remain within planetary boundaries as adoption scales.

\textit{Third}, future technologies must go beyond retrofitting the present; they must be fundamentally redesigned to work with, rather than against, biospheric constraints.The emerging field of ``living technologies'' highlights this direction and constitutes a growing research priority, although it still lacks rigorous definitional parameters for systematic scientific advancement. Some approaches directly harness biological processes, such as microbial fuel cells~\cite{Logan2006}, algae-based systems~\cite{Wijffels2010}, or engineered ecosystems. Others take inspiration from biology, including materials that self-repair or adapt, or modular architectures that tolerate slowness and local variation. 

The goal is not to mimic life's forms but to respect its operational envelope. Whether inspired by biological structures or developed independently~\cite{Fayemi2014_proceedings,Shimomura2012}, what unites these technologies is alignment with life’s physical logic: dilute energy fluxes, compatibility with ambient temperature and pressure, decentralized control, long feedback loops, evolutionary timescales, and circular material flows. These properties are not idealizations, but thermodynamic necessities. Technologies that follow them, whether directly biological, bio-inspired, or entirely novel, can gradually displace current zombie technologies and infrastructures that exceed planetary limits. In this sense, they are not only pathways to reduce harm but invitations to embed technology within regenerative ecological cycles.

The necessary conditions for such life-compatible systems are summarized in Fig.~\ref{fig:dual_pathways}B. Viable technologies must: (i) run on renewable fluxes; (ii) operate at low power density; (iii) rely on recyclable or abundant materials, with metals used at low concentrations; (iv) utilize biocompatible chemistry, typically based on CHNOPS elements; and (v) occupy land equitably without displacing ecological functions. Just as degrowth is a necessary (but not sufficient) systemic condition for sustainability, these design rules are necessary (but not sufficient) technological conditions: any design that fails even one is unlikely to be sustainable in the long run.

Such limits must inform not only individual behaviors or consumption choices, but also engineering, design, and infrastructure decisions. As technologies scale, the full arc of their operation becomes critical: where do materials and energy come from, how are they transformed, how long do they function, and what becomes of them at the end of life? Sustainability is not solely a question of operational efficiency; it is a question of systemic compatibility with the Earth’s geochemical and energetic boundaries. A device that is efficient and functions well in isolation may still be ecologically destabilizing if it draws on rare materials or produces unmanageable waste~\cite{Geyer2017}. While no material is strictly renewable, its use can approximate renewability if extraction, transformation, and disposal are designed to be circular, reversible, and non-destructive~\cite{Stahel2016,graedel2015basis}.

The ``energy transition'' must occur simultaneously in three dimensions: power (i.e., energy flow), the nature and quantity of materials (fewer minerals, genuine circularity), and the equitable sharing of land. Harvesting available renewable power on Earth (99\% of which originates from solar radiation) at high output generally requires far more materials and surface area than fossil fuels. For example, replacing global coal production ($\sim\!10^{5}$~kg~s$^{-1}$) with charcoal would require an area of $\sim\!10^{13}$~m$^2$, about one-third of all emerged land on Earth.

The chemistry of life-compatible technologies should favor abundant and biocompatible elements whose cycles are already embedded in the biosphere’s metabolism. More than the specific elements themselves (i.e., CHNOPS), what matters is whether the materials and reactions can be repeatedly integrated into Earth’s slow but stable biogeochemical cycles. Mechanical and chemical energy harnessed directly from renewable fluxes, rather than through combustion or energy-intensive conversion, often meets these criteria. In many cases, reuse outperforms recycling when judged by energy and material requirements~\cite{EEA2021}. Critically, sustainable technologies must operate slowly enough to allow feedback, adaptation, and repair—hallmarks of systems that persist over time.

To remain within planetary boundaries, future agricultural systems must radically reduce high-power operations while improving nitrogen and phosphorus use efficiency in both crop and livestock production. This imperative is even more acute in organic farming, where nutrient losses directly undermine yields~\cite{Chatzimpiros2023}. Ultimately, environmental nitrogen losses will depend not only on technological efficiency but also on total food demand and population scale.

Degrowth and life-compatible technologies are thus mutually reinforcing: the former reduces total throughput to within planetary boundaries, while the latter ensures that the remaining activity operates within the biosphere’s regenerative capacity.

\section*{Conclusion: The way forward}
\addcontentsline{toc}{section}{Conclusion}
\label{sec:conclusion}

The current technological ``Promethean'' paradigm~\cite{RizzoPetruccioli2023} is reaching its physical end. The crisis is systemic, encompassing climate, ecosystems, pollution, materials, energy, waste, and land use~\cite{Rockstrom2009,steffen2018trajectories}. It is not simply an energy crisis: although fossil reserves remain vast and the solar flux alone far exceeds human use, extraction, transformation, and deployment of resources at high speed, concentration and power inevitably generate waste and destabilize the ecosphere~\cite{Sassoon2009}.

Novel technologies and decarbonization will not cure the troubles created by previous waves of technology. Substituting fossil cars with electric vehicles or coal plants with solar farms may reduce emissions in the short term, but leaves untouched the deeper logic of expansion, acceleration, and material intensification. These are zombie technologies: operational yet fundamentally incompatible with long-term viability because they depend on finite stocks~\cite{graedel2015basis} and transgress environmental limits~\cite{Rockstrom2009}.

What lies ahead is a post-Promethean paradigm: one that accepts power descent, embraces slower and materially bounded processes, and redesigns ambition within thermodynamic realism. Longevity will come not from control but from coherence with planetary life. This requires reconceptualizing the ``energy transition'' not merely as a substitution of sources but as a systemic shift in both power and material use. We therefore call for interdisciplinary analyses that account for the full ecosphere: its resources, wastes, and boundaries across scales and criteria.

Long-term sustainability rests on two complementary transformations: degrowth~\cite{Hickel2020,kallis2018degrowth,Georgescu1971} and life-compatibility. Degrowth is multifactorial: it entails absolute reductions in extraction, throughput, and waste, along with decreases in the speed, size, and power of activities, guided by feedback regulation. Life-compatibility is likewise multifactorial: technologies must operate within the solar power budget, maximize recycling, avoid repeated dilution–concentration cycles, and rely on catalytic chemistry compatible with biogeochemical cycles~\cite{Stahel2016}.

These two paths are indissociable. Without life-compatible technology, degrowth lacks viability; without reductions in total throughput, even the most benign technologies scale into unsustainability. Integrated together, degrowth and life-compatibility reinforce one another, producing favorable feedbacks that make both feasible.

The natural sciences —physics, chemistry, biology, and geology— set the boundary conditions for human activity. They establish hard limits in principle, soft limits in practice, and environmental limits as planetary boundaries~\cite{Rockstrom2009,steffen2015,steffen2018trajectories,Richardson2023}. Degrowth is the only path compatible with these conditions, and its feasibility is evidenced by the persistence of life across billions of years~\cite{schramski2015human}. 

Political and economic choices must therefore be taken within these scientific constraints. “Green growth” remains speculative, lacking theoretical and empirical foundation~\cite{Parrique2019,hickelisgreengrowthpossible}. More credible options include modular and decentralized approaches to land allocation among humans, agriculture, wildlife, and machines. Targeted reductions in consumption —particularly those addressing disproportionate footprints— may prove the most viable route to stabilizing aggregate resource use while satisfying multiple sustainability constraints. Stabilizing material and power throughput inevitably carries major political implications, particularly for resource distribution mechanisms and socioeconomic restructuring~\cite{daly1977steady}.

Human obstacles, whether demographic, agricultural, social, political, economic, psychological, cultural, or military, make a degrowth transition to steady state difficult, but not impossible. Both current exponential growth and the illusion of green growth face inescapable constraints rooted in biophysical principles that cannot be violated. From a biophysical standpoint, a degrowth transition to steady state remains both the most realistic pathway and the best alignment with the constraints of planetary processes.

\subsection*{Acknowledgments} We thank Michael Gao, Ariane Guillemot, Mathieu Hernandez, and Roland Lehoucq for stimulating discussions and help with calculations. This work has benefited from a grant of the Agence Nationale de la Recherche under the France 2030 program (ANR-22-PERE-0003). Planet rendering in Fig. 1 adapted from Wikimedia Commons.


\bibliographystyle{unsrt}
\bibliography{sustainability.bib}

\cleardoublepage

\newpage

\beginsupplement

\begin{center}  

{\Large \bf The Physics of Sustainability: \\ Material and Power Constraints for the Long Term}
\medskip \medskip

{\Large by J. Halloy et al. }

\medskip \medskip \medskip \medskip \medskip \medskip \medskip \medskip \medskip \medskip

{\Large \bf Supplementary Material}  

\end{center} 

\medskip \medskip \medskip \medskip \medskip \medskip \medskip \medskip \medskip \medskip

\section{Data and figures} 
\label{sec:sassoon_figures}

The following data are from~\cite{Sassoon2009,Hermann2006,IEA2012,haff2014technology} expressed in TW ($10^{12}$~W). 
From an incident solar flux of 162 000 TW, the atmosphere absorbs energy (or destroys or converts exergy, or usable energy) at a rate of 31 000 TW, of which 870 TW appears as kinetic energy of the winds. The hydrosphere absorbs another 41 000 TW in evaporating water. Absorption by the biosphere of 15 000 TW leads via photosynthesis to the generation of chemical energy in plant matter at a rate of 90 TW, of which the technosphere appropriates almost 10 TW (agriculture and forestry). Fossil fuels, uranium and renewable energy sources provide energy to the technosphere at a rate of 17~TW.
This is an appreciable fraction of the geothermal energy flux (32 TW), the biochemical energy flux (90 TW) and the gravitational power load of the world's rivers (7 TW), suggesting the susceptibility of parts of the land surface, the biosphere and the fluvial portion of the hydrosphere to disruption or appropriation by the technosphere.

\section{Hurdles and false good ideas} 
\label{sec:hurdles}

For each technological idea, several questions should be asked~\cite{grille_analyse}: Is it based on a sound physical principle? If yes, can a prototype be built? If yes, can it be industrialised on a large scale? If yes, what are its side effects?

 For instance, for hydrogen, a confusion between three points should be avoided. Native hydrogen (for burning or to produce electricity) is a renewable flux, but is exists only in very specific geologic conditions, hence in localized places (e.g. Mali) and in small amounts. For batteries, it is only a storage (and currently extracted from hydrocarbons rather than water...), with many advantages, but is does not help much to provide energy, and it requires many materials. For fusion, it is not expected to have practical applications within a century, and the expenses in matter and energy are enormous, the wastes too; even a prototype such as ITER will not necessarily be operational (we do not know what materials could be used for ITER walls, able to resist the neutron flux without short term damages).
 
More generally, several seemingly clever technological ideas are either impractical or have a globally negative balance because the address only one specific question withour global reasoning (Table~\ref{table:false_good_ideas}).

\newpage

\section{Supplementary Tables}
\label{sec:supp_tables}

\begin{table}[ht]
\centering
\caption{Conversion factors used of energy and power comparisons}
\begin{tabular}{|l|r|}
\hline
\hline
{\it Prefix } & {\it Value} \\
\hline
kilo (k)  & 10$^3$ \\ 
\hline
mega (M)  & 10$^6$ \\ 
\hline
giga (G)  & 10$^9$ \\ 
\hline
 tera (T) & 10$^{12}$ \\ 
\hline
 peta (P) & 10$^{15}$ \\ 
\hline
 exa (E) & 10$^{18}$\\ 
\hline
 zetta (Z) & 10$^{21}$\\ 
\hline
 yotta (Y) & 10$^{24}$\\ 
\hline
\hline
{\it Time unit } & {\it Value in s} \\
\hline
min & 6 10$^1$ \\ 
\hline
hour & 3.6 10$^3$ \\ 
\hline
solar day & 8.6 10$^4$ \\ 
\hline
year & 3.2 10$^7$ \\ 
\hline
century & 3.2 10$^9$ \\ 
\hline
\hline
{\it Energy unit } & {\it Value in J} \\
\hline
kilowatt hour & 3.6 10$^6$ \\ 
\hline
watt year & 3.2 10$^7$ \\ 
\hline
tonne of oil equivalent  & 4.1 10$^{10}$ \\ 
\hline
barrel of oil equivalent  & 6.1 10$^9$ \\ 
\hline
calorie  & 4.2 \\ 
\hline
kilocalorie  (or Calorie) & 4.2 10$^3$ \\ 
\hline
ton of TNT  & 4.2 10$^9$ \\ 
\hline
electronvolt  & 1.6 10$^{-19}$ \\ 
\hline
joule per mole & 1.7 10$^{-24}$ \\
\hline
\hline
\end{tabular}
    \label{table:Babel_units}

\caption{The International System of Units expresses energies in joule (or kg m$^2$ s$^{-2}$), the work realised by a force of 1 newton over 1 meter. The watt (or kg m$^2$ s$^{-3}$) is the power of a joule per second. Several other units are used in everyday life, in media and in specific disciplines.
}
\end{table}

\begin{table}[ht]
\centering
\caption{Comparison of powers}
\begin{tabular}{|l|c|r|}
\hline
\hline
{\it Energy form } & {\it Upper limit} & {\it Current} \\
\hline
photovoltaic &  10$^{16}$ W & 10$^{11}$ W \\
\hline
hydroelectric & 10$^{12}$ W & 10$^{11}$ W \\
\hline
ocean thermal & 10$^{13}$ W & 10$^{6}$ W \\
\hline
geothermal & 10$^{13}$ W & 10$^{10}$ W \\
\hline
tide & 10$^{12}$ W & 10$^{9}$ W \\
\hline
native hydrogen & ? & 10$^{4}$ W \\
\hline
\hline
\end{tabular}
    \label{table:powers}

 \caption{All figures are rough orders of magnitude, expressed in W. Heating of Earth is of order of 10$^{14}$ W. World energy consumption 
 is of order of 10$^{13}$ W. Both are much smaller than the incident solar flux, of order of 10$^{17}$~W at the top of the atmosphere, 10$^{16}$~W at ground level.
  }
\end{table}

\begin{table}[ht]
\centering
\caption{Examples of false good ideas}
\begin{tabular}{|r|c|l|}
\hline
\hline
Situation to address & Proposed solution & Negative impact \\
\hline
\hline
decarbonate energy & nuclear electricity & produces wastes, non renewable\\
\hline
decarbonate energy &wind turbine & requires materials: magnet, concrete\\
\hline
decarbonate energy &  photovoltaic panel& matter, energy, water to produce it \\
\hline
air pollution in towns & electric transport & matter for battery,  \\
&& moves pollution to produce electricity\\
\hline
matter for glass production& glass recycling& similar energy cost\\
\hline
plastic pollution & wooden items & deforestation\\
\hline
uranium shortage& extract uranium from see water&plastic production and pollution \\
\hline
oil shortage & shale oil& water and soil pollution \\
\hline
greenhouse effect& CO$_2$ capture in wood& effect of limited duration, requires intrants \\
\hline
finding rare materials&space mining&huge cost in energy, impact on space \\
\hline
finding rare materials&deep ocean mining&huge cost in energy, impact on sea floor \\
\hline
greenhouse effect&geo-engineering of albedo& uncontrolled feedbacks on climate \\
\hline
coral reef degradation&coral reef 3D printing&huge amounts of plastic\\
\hline
\hline
\end{tabular}
    \label{table:false_good_ideas}
\end{table}

\clearpage
\newpage

\section{Supplementary Figures}
\label{sec:supp_figures}

\begin{figure}[h!]
    \centering
    \includegraphics[width=0.8\textwidth]{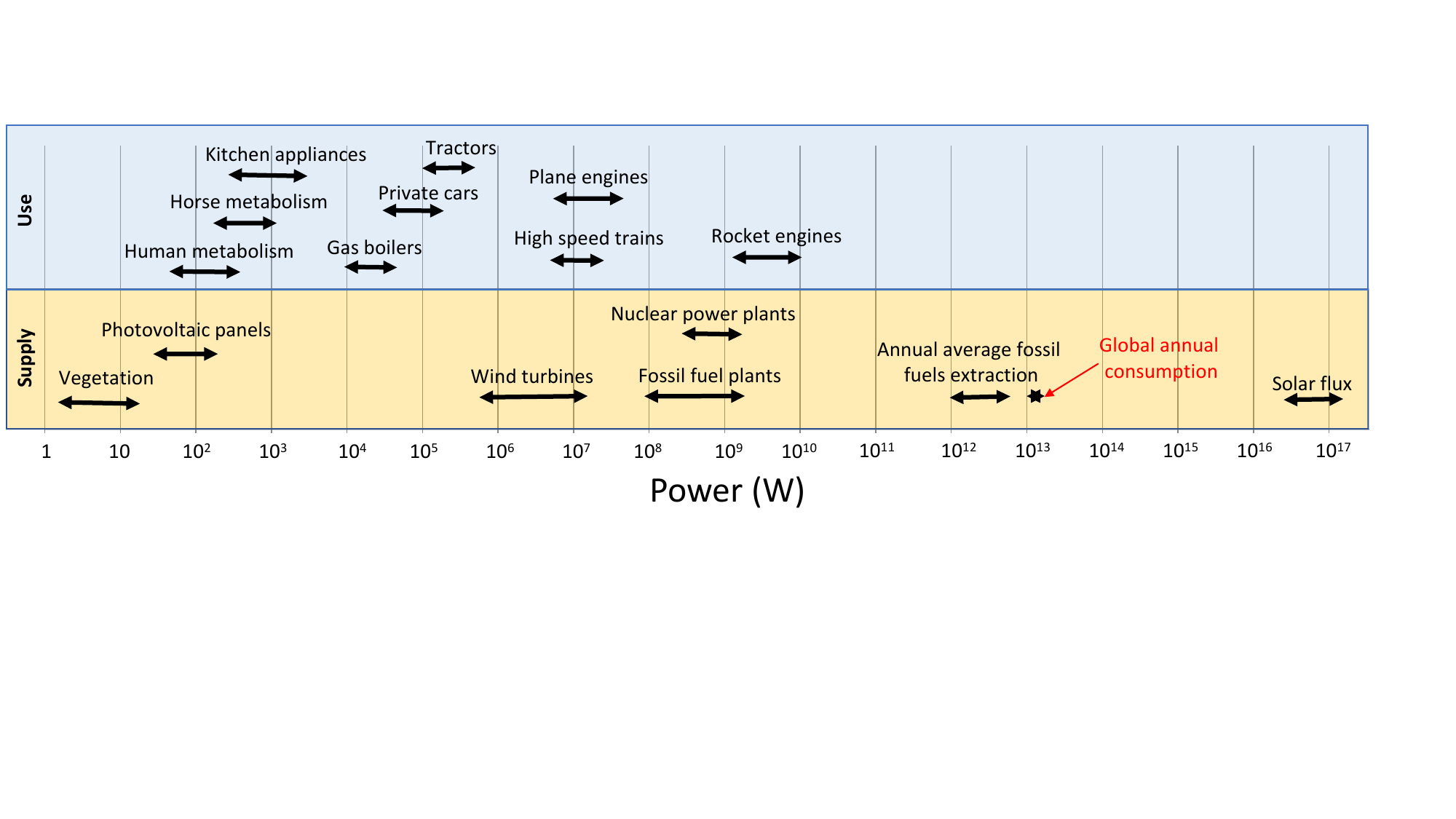}
    \vspace{-2.5cm}
	     \caption{%
	{\bf Power: orders of magnitude of energy converters.}
	Top: examples of human appliances. Bottom: examples of supply sources. All figures are in watt.
	}
	\label{fig:powers_and_fluxes}
	\label{fig:Powers_orders_of_magnitude_compared}
\end{figure}

\begin{figure}[t]
    \centering
   (A) \hfill   ~  \\ 
   \vspace{-4mm}
 \includegraphics[width=0.7\columnwidth]{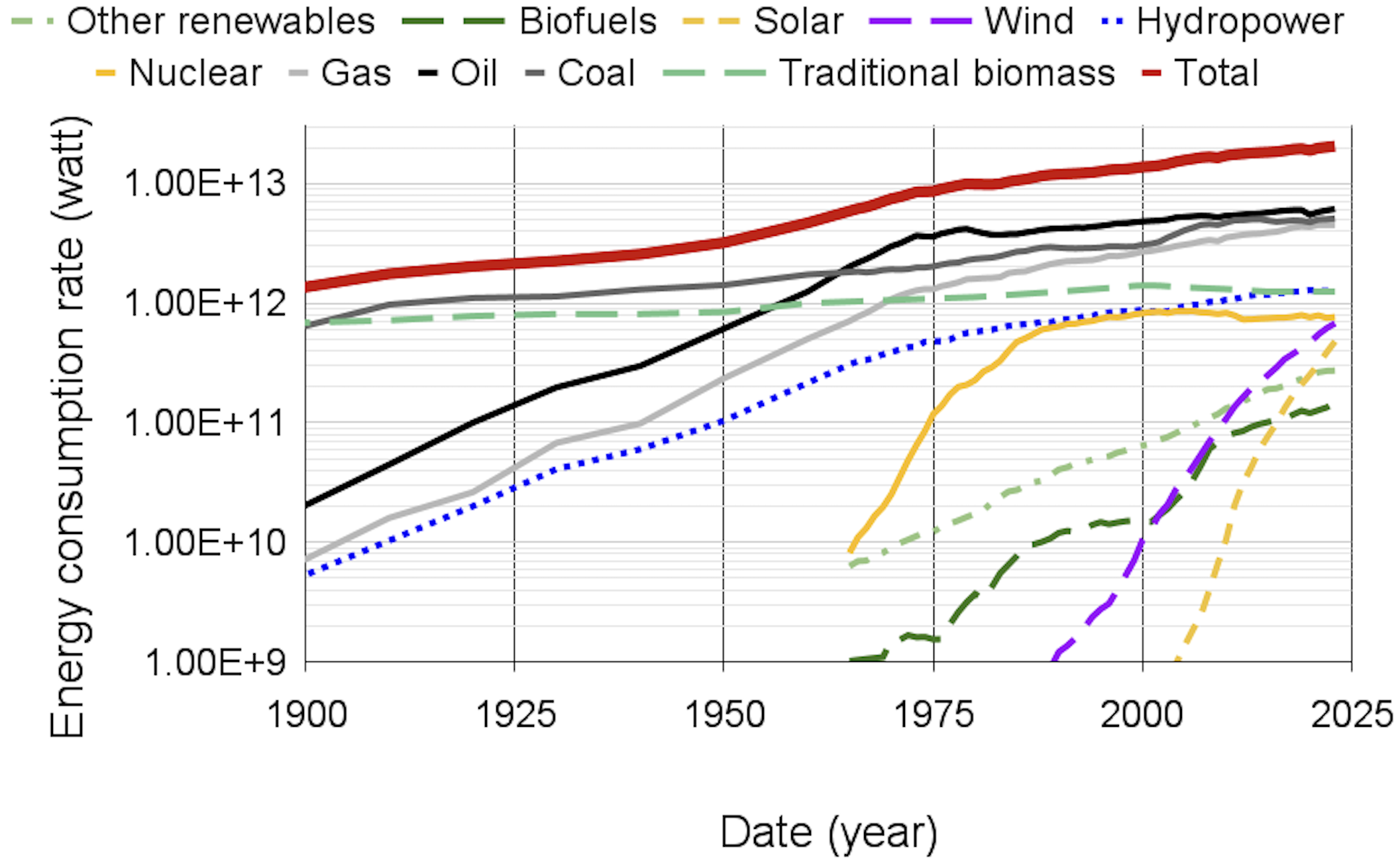}\\
   (B) \hfill   ~  \\ 
   \vspace{-4mm}
  \includegraphics[width=0.7\columnwidth]{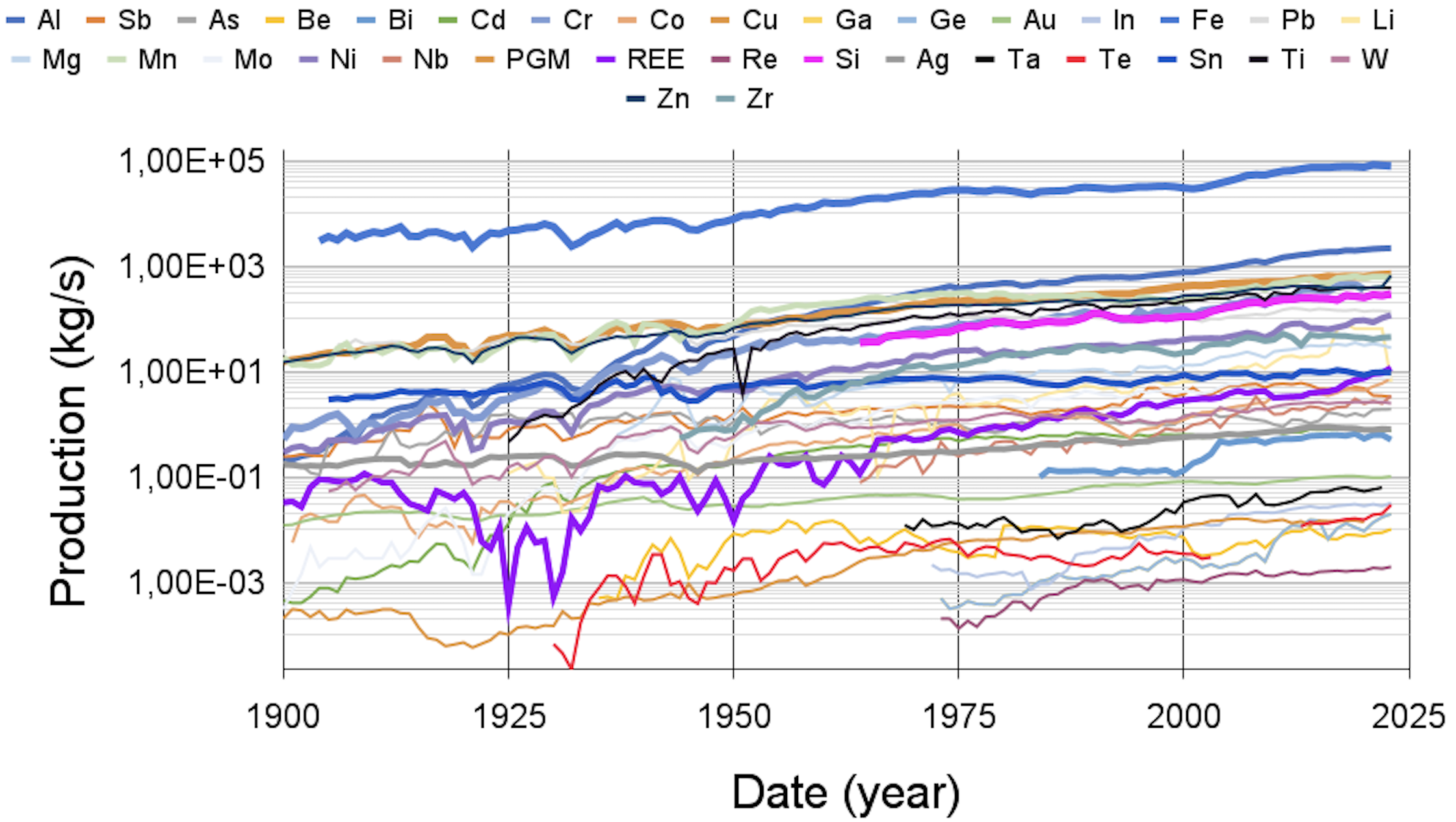}\\
  \hfill ~ \\
    \caption{%
   {\bf Increase of world resource consumption rates.} 
    {\bf(A)} Primary energy consumption rate vs date, in W  averaged over a one year sliding window.
Solid line: total. Dashes: fossil fuels extraction from stock, i.e. oil, gas (include shale), coal, uranium. Dots: extraction from flux, i.e. wind, hydraulic, solar.
   {\bf (B)} 
 Extraction rate of a few representative chemical elements.  ``PGM'': Platine Group Metals; ``REE'':
Rare Earth elements (lanthanides).
Note the semi-logarithmic scale.
   }
    \label{fig:increase_of_human_usages}
    \label{fig:Energy_consumption_vs_year}
    \label{fig: Material_extraction_vs_year}
\end{figure}

\begin{figure}[t]
    \includegraphics[width=0.8\textwidth]{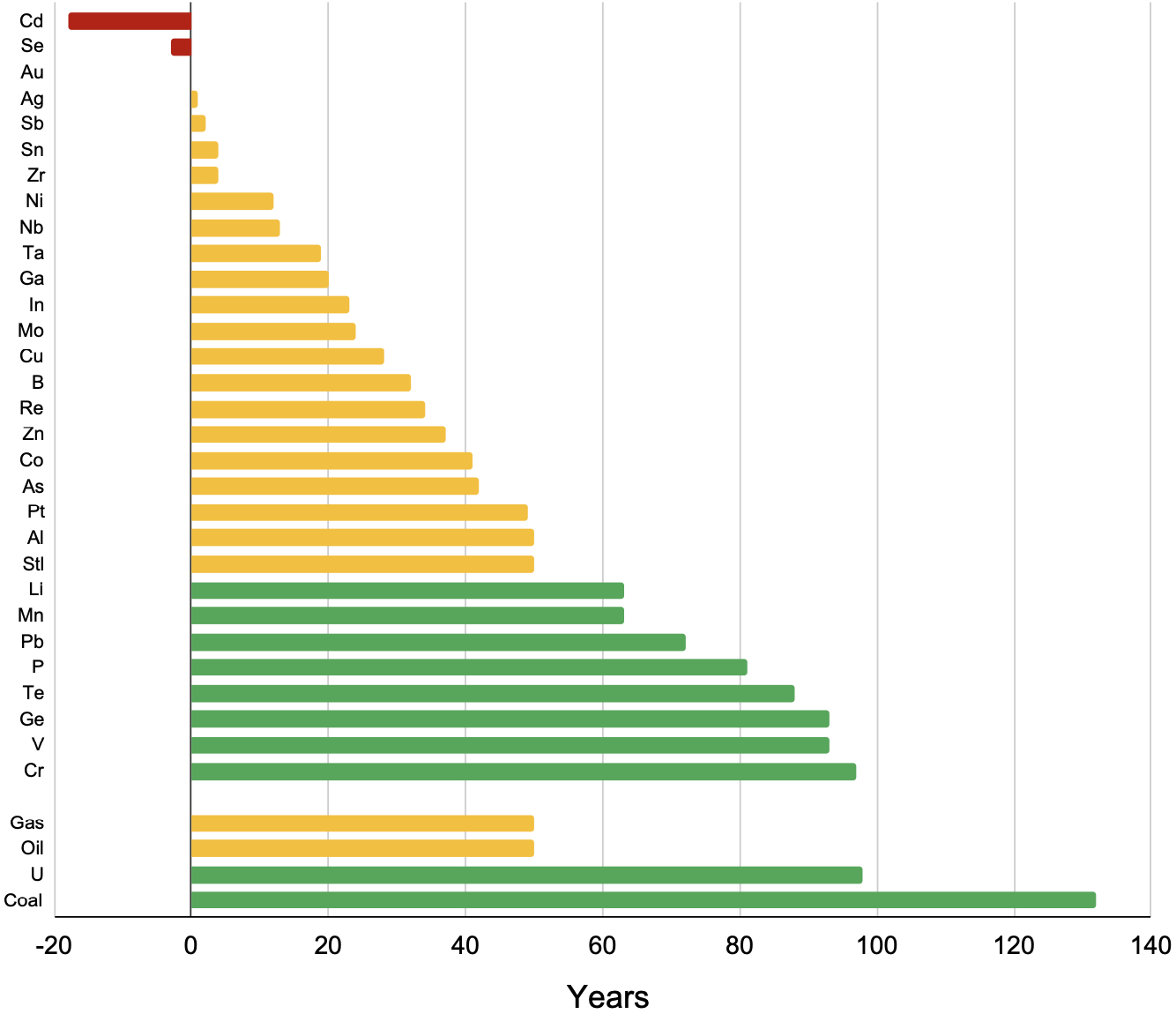}\\
   \caption{%
	{\bf Critical material and energy resources}
Same data as the material and energy peak year data of Fig.~\ref{fig:Rosetta_of_three_planetary_boundary_types}B, replotted linearly for legibility.
The abscissa indicates the year of peak. Red is for past peaks, yellow for peaks within half a century, green for peaks beyond that delay. 
Many material stocks (e.g. phosphorus, rare earth elements) are projected to peak within decades~\cite{heinberg2010peak,riondet2023applicability},  as will several metals~\cite{gordon2006metal}.
For instance,  copper  (Cu) could peak within 30-50 years~\cite{riondet2023applicability,northey2014modelling}.
Cadmium (Cd) and selenium (Se) have probably already peaked. 
For legibility, out of 92 elements, only a few  ones representative of the elements which peak within 200 years are plotted.
Data for ecosphere boundaries, which do not correspond to year predictions, are not plotted here.
    }
    \label{fig:Critical_materials_and_energy_vs_year}
\end{figure}

\end{document}